\documentclass[longbibliography,showpacs,floatfix,superscriptaddress,twocolumn,aps,prl]{revtex4-2}

\usepackage{floatrow}
\usepackage[caption=false]{subfig}
\usepackage{graphicx,color}
\usepackage{amsfonts}
\usepackage[figuresright]{rotating}  
\usepackage{amssymb}
\usepackage{bm}
\usepackage{amsmath}
\usepackage{mathtools}
\usepackage{psfrag}
\usepackage{multirow}
\usepackage{units}
\usepackage{lipsum}

\usepackage{physics}
\usepackage{soul}
\usepackage{titlesec}
\usepackage{times}
\usepackage[colorlinks=true,citecolor=blue,linkcolor=magenta,hypertexnames=false]{hyperref}
\usepackage{enumitem}
\usepackage[normalem]{ulem} 
\usepackage{cancel}

\usepackage[dvipsnames]{xcolor}
\usepackage[compat=1.1.0]{tikz-feynman}
\usepackage{tikz}
\usepackage{feynmp-auto}
\usepackage{tikz}
\usepackage{lipsum}
\usetikzlibrary{decorations.pathreplacing,decorations.markings,calc,intersections}
\newcommand {\beq} {\begin{equation}}
		\newcommand {\eeq} {\end{equation}}
	\newcommand {\bqa} {\begin{eqnarray}}
		\newcommand {\eqa} {\end{eqnarray}}
	\newcommand {\bseq} {\begin{subequations}}
		\newcommand {\eseq} {\end{subequations}}
	\newcommand {\baln} {\begin{align}}
		\newcommand {\ealn} {\end{align}}

	\renewcommand{\Tr}[1]{\mathrm{Tr}\left[#1\right]}

	\newcommand{\bk} {\textbf{k}}

	\newcommand{\om} {\omega}

		\newcommand{\tm} {T_{\text{cav}}}
			\newcommand{\dF} {\frac{\partial F}{\partial \omega}}
				\newcommand{\ddF} {\frac{\partial^2 F}{\partial \omega^2}}

\begin{document}

\title{Nonthermal electron-photon steady states in open cavity quantum materials}
\author{R. Flores-Calderón}
\affiliation{Max Planck Institute for the Physics of Complex Systems, Nöthnitzer Stra{\ss}e 38, 01187 Dresden, Germany}
\author{Md Mursalin Islam}
\affiliation{Max Planck Institute for the Physics of Complex Systems, Nöthnitzer Stra{\ss}e 38, 01187 Dresden, Germany}
\author{Michele Pini}
\affiliation{Max Planck Institute for the Physics of Complex Systems, Nöthnitzer Stra{\ss}e 38, 01187 Dresden, Germany}
\author{Francesco Piazza}

\affiliation{Theoretical Physics III, Center for Electronic Correlations and Magnetism,
Institute of Physics, University of Augsburg, 86135 Augsburg, Germany}
\affiliation{Max Planck Institute for the Physics of Complex Systems, Nöthnitzer Stra{\ss}e 38, 01187 Dresden, Germany}

\begin{abstract}

Coupling a system to two different baths can lead to novel phenomena escaping the constraints of thermal equilibrium. In quantum materials inside optical cavities, this feature can be exploited as electrons and cavity-photons are easily pulled away from their mutual equilibrium, even in the steady state. This offers new routes for a non-invasive control of material properties and functionalities. We show that the absence of thermal equilibrium between electrons and photons leads to reduced symmetries of the steady-state electronic distribution function. 
Moreover, by defining an effective temperature from the on-shell distribution function, we find a non-monotonic behaviour as a function of cavity frequency, consistent with recent experimental findings.
Finally, we show that, the non-thermal behaviour leads to qualitative modifications of the material’s properties, as the standard Sommerfeld expansion for observables is modified by a leading-order correction linearly proportional to the temperature difference between the two baths and to the frequency-derivative of the electron damping.
 \end{abstract}
\maketitle


\textit{Introduction}---  Out-of-equilibrium phenomena have resurfaced in multiple areas of physics as a way to circumvent the restrictions imposed by thermal equilibrium. Non-equilibrium steady states, resulting from the coupling of the system of interest to two different baths, are available in various configurations across multiple platforms, ranging from transport in condensed-matter systems, through lasers in atomic and solid-state systems, to active matter and biological systems. 
In transport for instance two thermal baths act as source and drain of electrons 
and induce a charge current \cite{vonKlitzing2020,PhysRevLett.128.175001,PhysRevLett.129.027402,PhysRevLett.129.166801}. In active matter such as molecular motors \cite{Needleman2017,Popkin2016, Xi2019, AM-nonequ,Pacchioni2016}, the degrees of freedom subject to energy input are different from the ones dissipating it. Similar situations arise in turbulence \cite{Pomeau2016,Lemoult2016,PhysRevLett.128.024502}, where energy is injected at large length scales and viscously dissipated at short length scales, or in a laser, where electrons inside atoms are externally excited while light is emitted through the mirrors of a cavity \cite{siegman1986lasers}.

\begin{figure}[ht!]
    \centering
     \sidesubfloat[]{\includegraphics[width= 0.9\columnwidth]{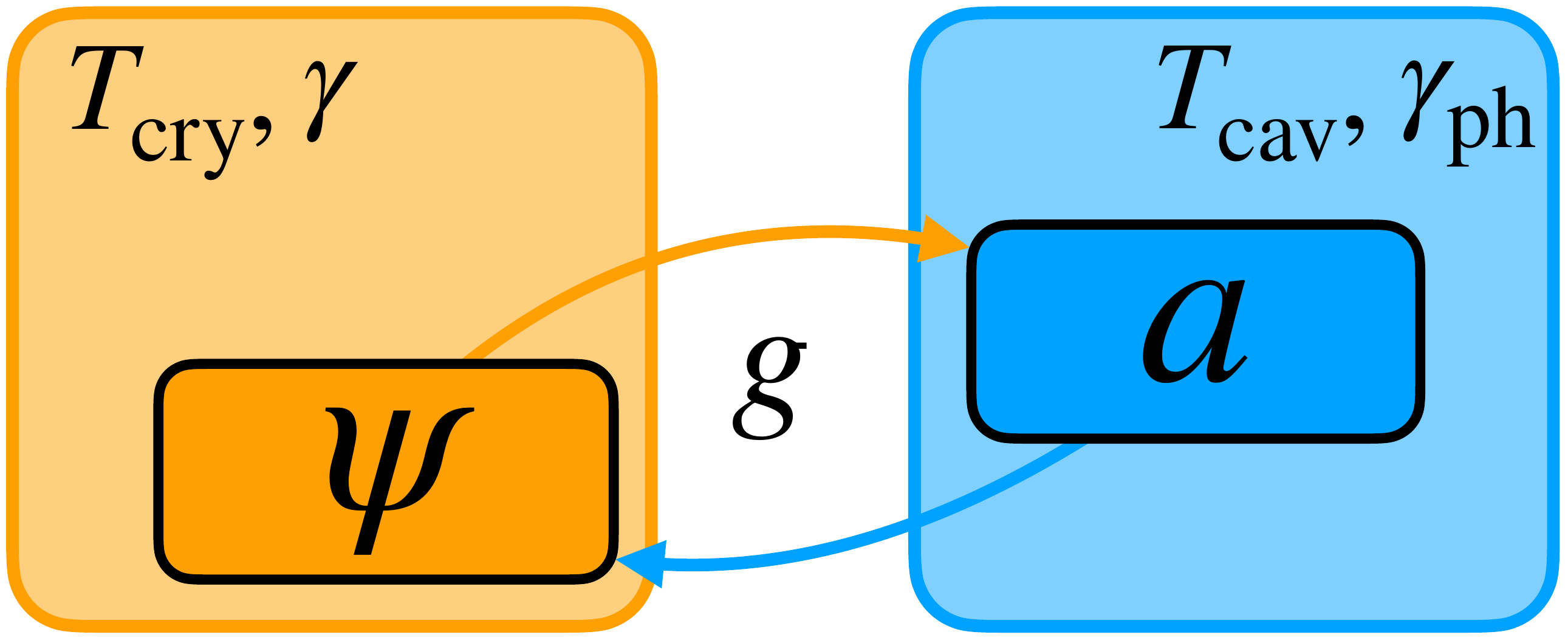}}\\
     \sidesubfloat[]{\includegraphics[width= 0.9 \columnwidth]{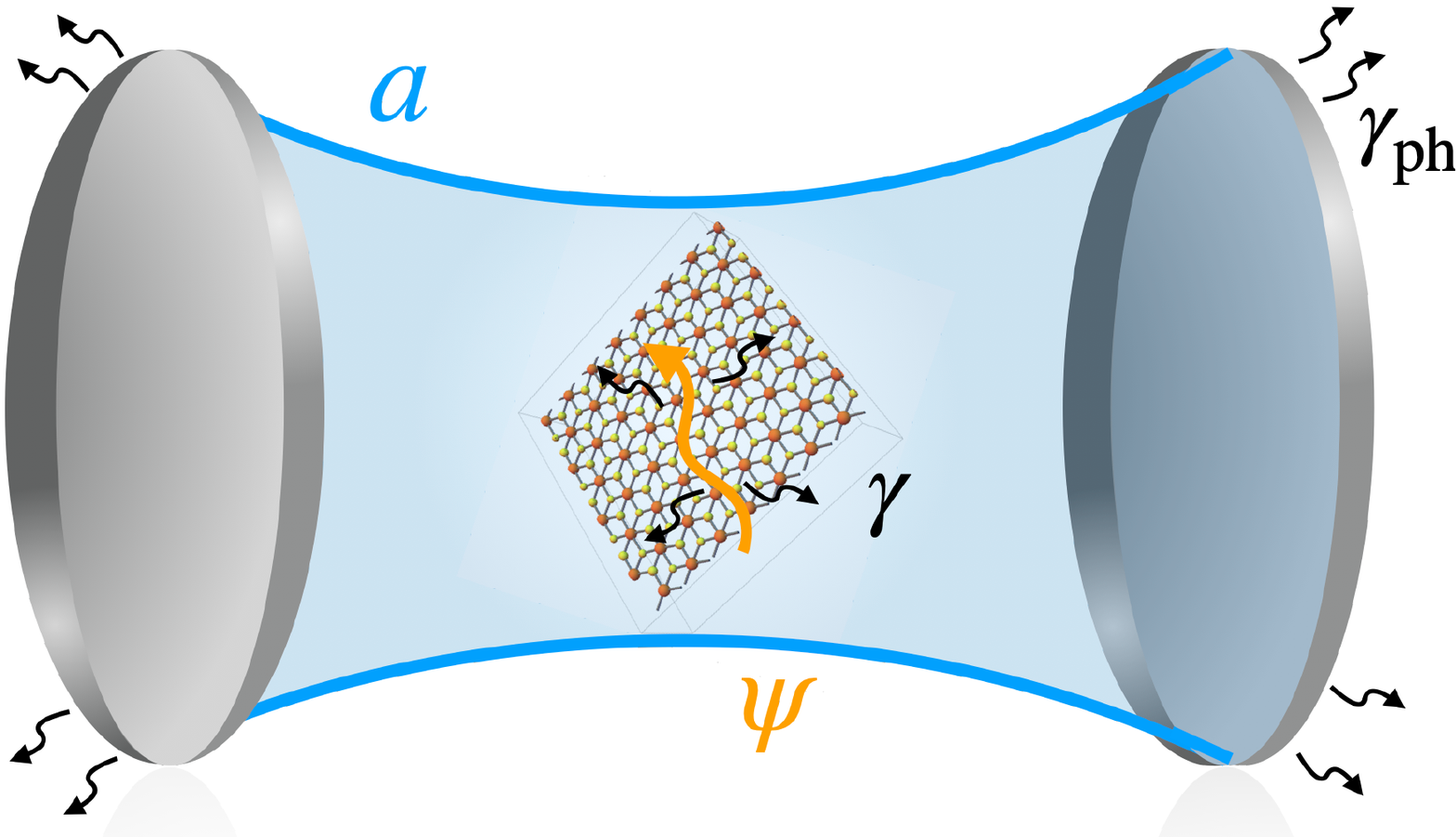}}
    \caption{ Schematic representation of the electron-photon system under consideration. (a) Two distinct thermal baths coupled to the electrons (photons)  $\psi$ ($a$) and with temperature $T_{\rm cry}$ ($T_{\text{cav}}$) and spectral width $\gamma$ ($\gamma_{\rm ph}$) in orange (blue). The electromagnetic field $a$ is coupled to electrons with a strength $g$. (b) Realization with a quantum material within a Fabry-Perot cavity.}
    \label{Fig1:Schematic}
\end{figure}

It is in this context that we turn towards cavity quantum materials \cite{garcia2021manipulating,mivehvar2021cavity,schlawin2022cavity,bloch2022strongly}, where electrons are coupled on the one hand to the cryostat directly attached to the material, and on the other hand to the discrete set of electromagnetic modes of the cavity, which in turn is coupled through the leaky mirrors to the continuum of electromagnetic modes outside, as schematically shown in Fig.~\ref{Fig1:Schematic}. Both the electromagnetic continuum and the cryostat act as thermal baths, but they do not need to be at the same temperature, so that a nonthermal steady state can be achieved.
Confining light around quantum materials through cavities has recently emerged as an alternative to the laser-based control \cite{de2021colloquium}, with the advantage that weakly, thermally excited electromagnetic fields can be used to affect the material without the large energy input restricting laser-based approaches to pulsed (transient) regimes. In particular, the absence of thermal equilibrium between electrons and photons has been identified as a source of novel phenomenology and enhanced control \cite{curtis2019cavity,PhysRevLett.127.177002,chakraborty2022controlling,eckhardt2024theory,vinas2023controlling,1T-TaS2-Exp}.


\floatsetup[figure]{style=plain,subcapbesideposition=top}

The present work is particularly motivated by a recent experiment demonstrating
cavity control of the metal-to-insulator transition in 1T-$\text{TaS}_2$ \cite{1T-TaS2-Exp}. The critical temperature associated with the charge-density-wave formation could be substantially modified by tuning only the cavity resonant frequency, despite the light-matter coupling being small relative to the intrinsic electronic scales. 
So far, the cavity-induced thermal Purcell effect has been suggested as a possible explanation \cite{1T-TaS2-Exp,chiriaco2023thermal}, whereby the electrons, due to the cavity environment, experience a temperature different from the one of the cryostat.

In this work, we instead focus on the non-thermal nature of the steady-state of electrons. We show that, the electronic distribution function has  a reduced symmetry compared to the thermal equilibrium one, parametrically tuned by the electron-photon coupling, quasiparticle damping and cavity frequency. From the on-shell nonthermal distribution, a low-energy effective temperature can be extracted from the vicinity of the Fermi surface which presents a non-monotonic behaviour as a function of the cavity frequency, in line with the non-monotonic behavior of the critical temperature observed experimentally in Ref.~\cite{1T-TaS2-Exp}. Moreover, we show that the dominant fluctuation effect on observables is genuinely nonthermal, as the standard Sommerfeld expansion is modified by a correction linearly proportional to the temperature difference between the cavity photons and the cryostat, as well as to the frequency-derivative of the electron damping. In this first work, we do not consider the specific charge-density-wave scenario of \cite{1T-TaS2-Exp}, but rather a simpler model of a two-dimensional metal inside a Fabry-Perot cavity, which makes the generic nature of the proposed nonthermal mechanism clear. We finally note that the nonthermal nature of the electronic distribution induced by cavity-photons has been already investigated in terms of its effect on the superconducting gap \cite{curtis2019cavity}, in analogy to the original Eliashberg effect with oscillating radiofrequency fields \cite{eliashberg1970film}. Here, we focus instead on general effects that the cavity-induced nonthermal behaviour has on the material's properties, namely how the fermionic distribution is modified, and how this influences the Sommerfield expansion for single-particle observables.


 \textit{Model}---
We treat the cavity as two perfectly conducting parallel mirrors with the electrons moving in a plane parallel to the mirrors and placed amidst them, as schematically shown in Fig.~\ref{Fig1:Schematic}(b). Photons inside the cavity acquire a finite mass because of the boundary conditions, which is set by the cavity fundamental frequency $\nu_0$, appearing in the photon dispersion $\nu_{\textbf{q}}^2=\nu_0^2+c^2\textbf{q}^2$. For simplicity, we will set $c=1$ and $\hbar=1$ throughout the paper. The Hamiltonian for the uncoupled, closed photon-electron system is given by:
\begin{align}
    H_0 = \sum_{\textbf{k}} \epsilon_{\textbf{k}}\psi^\dagger_{\textbf{k}}\psi_{\textbf{k}}+\sum_{\textbf{q}} \nu_{\textbf{q}}\left(a^\dagger_{\textbf{q}}a_{\textbf{q}}+\dfrac{1}{2}\right),
\end{align}
where the first term models the metallic behaviour of the electrons with a gapless dispersion $\epsilon_{\textbf{k}}= \lvert\textbf{k}\rvert^2/(2m)$ in $d+1$ space-time dimensions. We have expressed the Hamiltonian in terms of the creation and annihilation operators for the electrons $\{\psi_{\textbf{k}},\psi^\dagger_{\textbf{k}'}\}=\delta_{\textbf{k},\textbf{k}'}$, as well as for the cavity photons $[a_{\textbf{q}},a^\dagger_{\textbf{q}'}]=\delta_{\textbf{q},\textbf{q}'}$. We couple the electrons to a thermal bath at temperature $T_{\textrm{cry}}$, which physically corresponds to a cryogenically cooled substrate.
We choose the simplest model where the bath can be treated exactly. It is composed of fermionic degrees of freedom coupled linearly to the electrons:
\begin{align}
\label{eq:e-bath}
    H_{\text{e-bath}} = \sum_{\textbf{k},s} \big[\epsilon_s(\textbf{k})f^\dagger_{s\textbf{k}}f_{s\textbf{k}}+t_s(\textbf{k}) (\psi^\dagger_{\textbf{k}}f_{s\textbf{k}}+f^\dagger_{s\textbf{k}}\psi_{\textbf{k}})\big].
\end{align}
 Here we have an extensive number of bath degrees of freedom $f_{s\textbf{k}}$, labeled by $s$ and coupled to each electron momentum component. As shown in the Supplementary Material, where we adopt a real-time path-integral formulation on the Keldsyh contour \cite{kamenev2023field,Keldysh-OQS} to integrate the bath out, the latter enters the effective theory for the electrons via its temperature $T_{\textrm{cry}}$ and the quasiparticle damping $\gamma(\textbf{k},\omega)=\frac{\pi}{2} \sum_s t_s^2(\textbf{k}) \ \delta(\omega-\epsilon_s(\textbf{k}))$. A description in terms of these two quantities is actually more generally applicable than the simple model of Eq.~\eqref{eq:e-bath}.
 Additionally, the leaky cavity mirrors allow the external environmental photons at temperature $T_{\textrm{cav}}$ to act as a thermal bath for the cavity modes. The Hamiltonian for the photon-bath and its coupling to the cavity photons reads:
\begin{align}
    H_{\rm ph-bath}= &\sum_{s,\textbf{q}} \bigg[\nu_s(\textbf{q})\left(d^\dagger_{s\textbf{q}}d_{s\textbf{q}}+\dfrac{1}{2}\right)\notag \\&+\frac{t^{\rm ph}_s(\textbf{q})}{2\sqrt{\nu_{\mathbf{q}} \nu_s(\mathbf{q})}}(a^\dagger_{-\textbf{q}}+a_{\textbf{q}})(d^\dagger_{s-\textbf{q}}+d_{s\textbf{q}})\bigg],
\end{align}
where we again consider an extensive number of harmonic oscillators $d_{s\textbf{q}}$ coupling to each photonic mode of the cavity in a linear fashion. Similarly to the electrons, we integrate out the photon-bath within our path-integral formulation to obtain an effective description in terms of the bath temperature $T_{\text{cav}}$ and the resulting photon damping $\gamma_{\rm ph}(\textbf{q},\omega)= \sum_s \frac{\pi t^{\rm ph}_s(\textbf{q})^2}{4\nu_s(\textbf{q})}\delta(\omega-\nu_s(\textbf{q}))$. Finally, we model the electron-photon coupling by a Yukawa-type interaction:
\begin{align}
    V= g \int \dd^d\bm{x} \ \phi(\bm{x}) \psi^\dagger(\bm{x})\psi(\bm{x}),
    \label{Yukawa-coupling}
\end{align}
where $\psi(\bm{x})$ and $\phi(\bm{x})$ are the fermionic and real bosonic Fourier transformed operators of $\psi_{\textbf{k}}$ and $\phi_\textbf{q}=(a_{\textbf{q}}+a^\dagger_{-\textbf{q}})/\sqrt{2L\nu_{\textbf{q}}}$ respectively, 
$g$ is the light-matter coupling and $L$ is the cavity length. Throughout the rest of the paper, we will set $L$ by its relation with the fundamental cavity frequency $L=\pi/\nu_0$. In Fabry-Perot cavities $g$ is small, typically of the order of $0.1$ of the electron bandwidth \cite{schlawin2022cavity,Cavalleri-SC,andolina2024can}. Note that we are using a coupling to the electron density. This might be appropriate for deep-subwavelength cavities \cite{andolina2024can}, but for the Fabry-Perot cavity considered here, a current coupling should actually be used. We argue however that the qualitative non-equilibrium features of the system can be highlighted also with the simpler density coupling of Eq.~(\ref{Yukawa-coupling}). For the sake of simplicity, we also consider a momentum-independent coupling $g$. The total Hamiltonian we consider is thus $H=H_{0}+H_{\rm e-bath}+H_{\rm ph-bath}+V$.

  \textit{Dyson equation approach}---
  Our goal is to compute the electron distribution function $F(\textbf{k},\omega)$ in the steady state of the open system.
  We cannot assume thermal equilibrium, plus electrons and photons cannot in principle be treated classically. 
  Therefore, a classical (Langevin or Fokker-Planck) approach is not suitable, while a standard Boltzmann-type equation for the electrons is restricted to the case of a small electron-bath coupling ensuring well-defined quasiparticles \cite{curtis2019cavity}. We thus choose to start with coupled Dyson equations for the electron's and photon's distribution functions. Since we are dealing with an interacting system -- the electron-photon coupling is not linear -- we adopt a self-consistent one-loop approach \cite{piazza2014quantum,Rao-Francesco} which can be obtained within the real-time path-integral formulation illustrated in the Supplementary Material. We proceed now by neglecting the back action of the electrons onto the photons, which is valid as long as the photon-bath coupling, quantified by $\gamma_{\rm ph}$, is sufficiently larger than the coupling $g$. 
 Further assuming space- and time-translation invariance in the steady state, the resulting equation for the electron distribution function reads:
\begin{equation}
    \begin{split}
        &\gamma(\textbf{k},\omega) \Delta F(\textbf{k},\omega)=g^2 \sum_{\textbf{k}'}\int_{\tilde{\omega}} A(\textbf{k}',\tilde{\omega})A_{\rm ph}(\textbf{k}'-\textbf{k},\tilde{\omega}-\omega)    \\
        &\left\{ B_0(\tilde{\omega}-\omega)\big[F(\textbf{k}',\tilde{\omega})-F(\textbf{k},\omega)\big]-1+F(\textbf{k},\omega)F(\textbf{k}',\tilde{\omega}) \right\},
    \end{split}
    \label{QKE}
\end{equation}
where $\int_{\tilde{\omega}}\equiv\int\frac{\dd \tilde{\omega}}{2\pi} $, and we have defined the deviation of the distribution function from equilibrium $ \Delta F(\textbf{k},\omega) = F(\textbf{k},\omega)-F_0(\omega)$,  along with the thermal distribution functions for the electrons $F_0(\omega)=\tanh{[(\omega-\mu_0)/(2T_{\text{cry}})]}$ and the photons $B_0(\omega)=\coth[\omega/(2T_{\text{cav}})]$. We have introduced here the electron chemical potential or Fermi energy $\mu_0$ set by the bath. We have also written the right hand side in terms of the spectral functions of the electrons $A(\textbf{k},\omega)$ and the photons $A_{\rm ph}(\textbf{k},\omega)$, which 
are defined as $A(\textbf{k},\omega)=\frac{\gamma(\textbf{k},\omega)}{(\omega-\epsilon_{\textbf{k}})^2+  \gamma^2(\textbf{k},\omega)}$ and $\ A_{\rm ph}(\textbf{q},\omega)= \frac{\nu_0}{\pi}\frac{\gamma_{\rm ph}(\omega)}{(\omega^2-\nu_{\textbf{q}}^2)^2+  \gamma_{\rm ph}^2(\omega)}$.  Within the simple model introduced above, the quasiparticle dampings $\gamma$ and $\gamma_{\rm ph}$ are set by the baths.

A further approximation can be done by assuming that the photon spectral function varies in frequency on a much smaller scale than the electron spectral function and the distribution functions of both photons and electrons. In order for this to be true, the photon damping, setting the width of the spectral function, must be sufficiently smaller than the electron damping, $\gamma_{\rm ph}\ll\gamma$, as well as the temperatures, $\gamma_{\rm ph}\ll T_{\rm cry},T_{\rm cav}$.
Under these conditions, we can treat the photon spectral function $A_{\text{ph}}(\textbf{k},\omega)$ as a Dirac-delta peaked at the photon dispersion to perform the frequency integral in Eq.~(\ref{QKE}). Furthermore, the momentum integral can be similarly performed by observing that the photons in the cavity have an extremely light mass compared to the electronic effective mass coming from the dispersion relation in typical solid-state materials. This allows to treat the momentum-dependence of the photon spectral function also as a delta function at zero momentum. As shown in the Supplementary Material, performing these approximations leads to a self-consistent equation for the distribution function of the form:
\begin{align}
\begin{split}
    \Delta F(\textbf{k},\omega) \ = \ \dfrac{ g^2 }{4\pi \gamma(\textbf{k},\omega)}&\big[A(\textbf{k},\omega+\nu_0)H_{\nu_0}(\textbf{k},\omega)\\
    &-A(\textbf{k},\omega-\nu_0)H_{-\nu_0}(\textbf{k},\omega)\big],\label{F_QKE}
\end{split}
\end{align}
where we defined 
$H_{\nu}(\textbf{k},\omega)=B_0(\nu)\big[F(\textbf{k},\omega
+\nu)-F(\textbf{k},\omega)\big]+F(\textbf{k},\omega)F(\textbf{k},\omega+\nu)-1$.  Eq.~\eqref{F_QKE} can be solved iteratively, starting from the thermal distribution $F_0(\omega)$. Before discussing the full numerical solution, let us consider the following two regimes: $\nu_0\ll T_{\text{cav}},T_{\text{cry}},\gamma$ or $\nu_0\gg T_{\text{cav}},T_{\text{cry}},\gamma$.  In the Supplementary Material, we show that, at the lowest order in both regimes, one finds $F(\textbf{k},\omega)=F_0(\omega)$. 
Moreover, in the small $\nu_0$ limit,  Eq.~(\ref{F_QKE}) can be expanded to the linear order in $\nu_0$ and assumes the form of a non-linear differential equation, for which a further analytic treatment is possible. Given a $g$ which is much smaller than the electronic energy scales, we can perform a weak coupling expansion of the non-linear differential equation (see Supplementary Material), which results in the leading order correction for the distribution:
\begin{equation}
    \begin{split}
    &\Delta F(\textbf{k},\omega)  \\
    &=\dfrac{g^2 \nu_0\Delta T \gamma(\textbf{k},\omega)}{4\pi T_{\text{cry}} A(\textbf{k},\omega)}\dfrac{\partial}{\partial \omega}\left[ \dfrac{A(\textbf{k},\omega)}{\gamma(\textbf{k},\omega)} \ \text{sech}\left(\frac{\omega -\mu _0}{
  2 T_{\text{cry}}}\right)\right]^2,
  \end{split}
  \label{F_sol_approx}
\end{equation}
where the temperature difference is defined as $\Delta T = T_{\text{cav}}-T_{\text{cry}}$. In this case, the out of equilibrium contribution is proportional to the temperature difference between the two baths, thus $\Delta T$ amplifies the photon-induced redistribution. 
In Eq.~(\ref{F_sol_approx}) we observe an explicit momentum dependence of the nonthermal distribution, which is absent in the thermal distribution $F_0(\omega)$. Moreover, we see that the nonthermal deviation $\Delta F(\textbf{k},\omega)$ is parametrically tuned by the coupling $g$, the quasi-particle damping $\gamma(\textbf{k},\omega)$ and the cavity frequency $\nu_0$. As we will show later, this analytical solution is also able to capture the leading-order behavior of the on-shell effective temperature for small $\nu_0$ and $g$.

\begin{figure}[tb!]
 \includegraphics[width= \columnwidth]{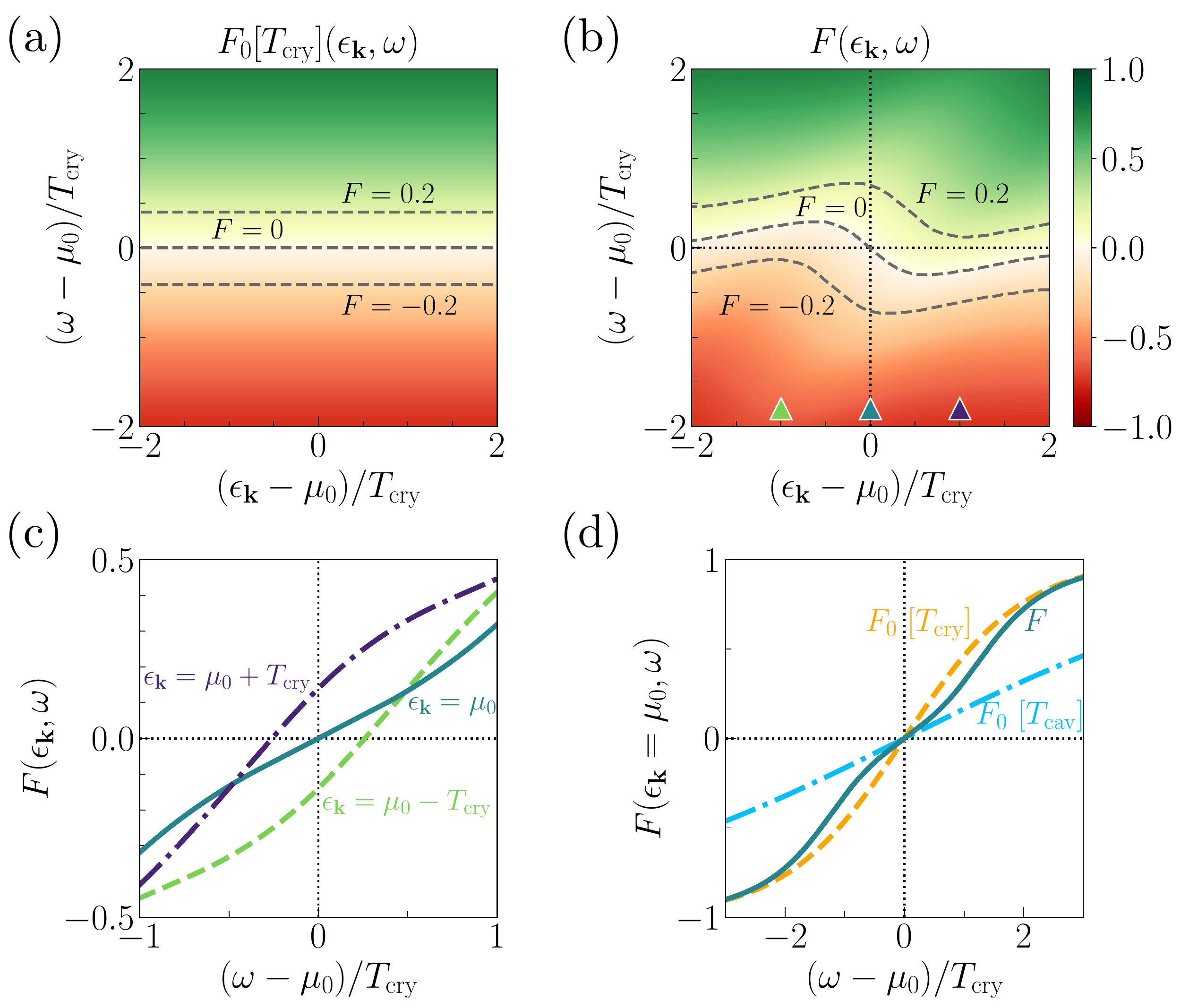}
    \caption{ Distribution function from the numerical solution of Eq.~\eqref{F_QKE}. (a) Non-interacting thermal solution $F_0(\omega)$ at the cryostat temperature $T_{\text{cry}}$. This is anti-symmetric around the line $\omega=\mu_0$. (b) Nonthermal distribution $F(\textbf{k},\omega)$ with fixed 
    $T_{\text{cav}}/T_{\text{cry}}=3$, $\nu_0/T_{\text{cry}}=0.5$, $\gamma_0/T_{\text{cry}}=0.5$, $\mu_0/T_{\text{cry}}=100$, $g/T_{\text{cry}}=1$. Note that it has anti-inversion symmetry about the point $(\epsilon_{\textbf{k}}=\mu_0,\omega=\mu_0)$. (c) Nonthermal distribution cuts for different momentum values  (fixed through $\epsilon_{\textbf{k}}$), corresponding to the colored triangles in panel (b). (d) Comparison of the thermal distributions $F_0[T](\omega)=\tanh[(\omega-\mu_0)/(2T)]$ at $T=T_{\text{cry}},T_{\text{cav}}$ and the nonthermal $F(\textbf{k},\omega)$ at the Fermi momentum ($\epsilon_{\textbf{k}}=\mu_0$).}   
    \label{fig:2}
\end{figure}

 \textit{Broken thermal symmetry and steady-state distribution function}---
 We turn now to discuss the full solution of Eq.~\eqref{F_QKE}. For $g \rightarrow \infty $, the equation is dominated by the right hand side and the solution is the Fermi-Dirac distribution with the photon-bath temperature $T_{\text{cav}}$, as proved in the Supplementary Material. This is natural since in this limit the coupling to the photons (and their respective bath) dominates. In the opposite limit $g\rightarrow 0$, the right hand side in Eq.~\eqref{F_QKE}  vanishes, such that the solution becomes $F(\textbf{k},\omega)=F_0(\omega)$, i.e.~the Fermi-Dirac distribution function at temperature $T_{\text{cry}}$, which is trivially expected for a vanishing coupling to the photons. 
 For generic values of $g$ the electron distribution is instead nonthermal. We focus here on the simple case of a momentum and frequency-independent electron dissipation $\gamma_0=\gamma(\textbf{k}, \omega)$, and we show the corresponding $F(\textbf{k},\omega)$ in Fig.~\ref{fig:2}. Note that, for a momentum-independent dissipation $\gamma_0$, the momentum dependence of $F(\textbf{k},\omega)$ enters only through $\epsilon_{\textbf k}$, i.e.~$F(\textbf{k},\omega)=F(\epsilon_{\textbf k},\omega)$. The nonthermal nature of the distribution becomes clearly visible when comparing the upper two panels. Thermal symmetry for $g\rightarrow 0$ fixes the distribution function $F(\textbf{k},\omega)=F_0(\omega)$ to be momentum independent, which means that in the $(\epsilon_{\textbf{k}},\omega)$ plane we have an inversion anti-symmetry along the $\omega=\mu_0$ line given by $F_0(\mu_0+\chi) =-F_0(\mu_0-\chi)$,  as shown in Fig.~\ref{fig:2}(a). For finite coupling, the distribution function develops a momentum dependence and the anti-symmetry around $\omega=\mu_0$ is broken. Instead, the distribution has now an anti-symmetry around the inversion point $(\epsilon_{\textbf{k}},\omega)=(\mu_0,\mu_0)$ given by $F(\mu_0+\xi,\mu_0+\chi)= -F(\mu_0-\xi,\mu_0-\chi)$, as easily seen in the contours shown in Fig.~\ref{fig:2}(b). We remark here that the reduced symmetry could further be completely broken with the inclusion of a frequency dependent damping $\gamma(\omega)$. Furthermore, the frequency dependence of the distribution is shown in Fig.~\ref{fig:2}(c), where we plot $F(\epsilon_{\textbf k},\omega)$ at fixed values of $\epsilon_{\textbf k}$. We see that away from the Fermi momentum ($\epsilon_{\textbf k} \neq \mu_0$), the zero crossing of $F$ is shifted from the origin, which also indicates a nonthermal nature. Even for the distribution at the Fermi momentum ($\epsilon_{\textbf{k}}=\mu_0$), we see a nonthermal interpolation of $F(\mu_0,\omega)$ between the distribution at the cavity temperature  around the Fermi energy $\omega\approx \mu_0$ and the distribution at the cryostat temperature at large frequencies, as depicted in Fig.~\ref{fig:2}(d).

\begin{figure}[t]
 \includegraphics[width= 0.9\columnwidth]{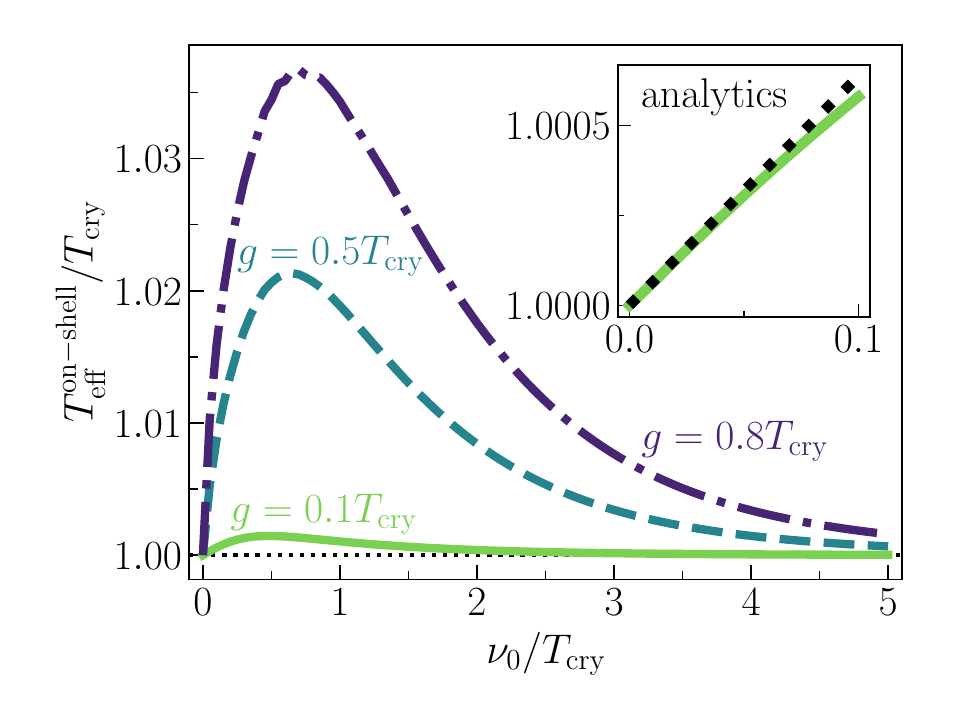}
 \caption{Effective on-shell temperature $T_{\text{eff}}^{\text{on-shell}}=\frac{1}{2}\big[ \frac{\partial F(\textbf{k},\epsilon_{\textbf{k}})}{\partial \epsilon_{\textbf{k}}}\big]^{-1}_{\epsilon_{\textbf{k}}=\mu_0}$ calculated from the numerical solution of Eq.~(\ref{F_QKE}) vs cavity  frequency $\nu_0$, for three values of the coupling $g$ and  with parameters $T_{\text{cav}}/T_{\text{cry}}=3$, $\mu_0/T_{\text{cry}}=100$ and  $\gamma_0/T_{\text{cry}}=0.5$. The inset shows the agreement of the full numerical solution (green solid line) with the analytical solution $T_{\text{eff}}^{\text{on-shell}}=(1-\alpha)^{-1}T_{\text{cry}}, \quad \alpha=\frac{ g^2\nu_0\Delta T }{4\pi \gamma_0^2T_{\text{cry}}^2}$ from Eq.~(\ref{F_sol_approx}) (dotted black line) for small $\nu_0$ and $g$.}
 \label{fig:3}
 \end{figure}

\textit{Effective on-shell temperature in the nonthermal steady state}---
Although the electron distribution is nonthermal, it is useful to look at its low-energy properties near the Fermi surface. Like in the previous section, we simplify our treatment and assume $\gamma(\textbf{k},\omega)=\gamma_0$. We can extract an effective temperature $T_{\text{eff}}^{\text{on-shell}}$ from the on-shell nonthermal distribution $F(\textbf{k},\omega=\epsilon_\textbf{k})$ by linearizing it around $\epsilon_\textbf{k}\simeq \mu_0$, i.e.~$T_{\text{eff}}^{\text{on-shell}}=\frac{1}{2}\big[ \frac{\partial F(\textbf{k},\epsilon_{\textbf{k}})}{\partial \epsilon_{\textbf{k}}}\big]^{-1}_{\epsilon_{\textbf{k}}=\mu_0}$. 
In a similar spirit to the thermal Purcell effect \cite{1T-TaS2-Exp,chiriaco2023thermal}, we can thus observe that the effective electron temperature depends on the cavity geometry, which enters in our model through the cavity frequency $\nu_0$. We remark however that here, in contrast to the thermal Purcell picture, the electrons are in a \emph{nonthermal} distribution $F(\textbf{k},\omega)$, and the effective temperature $T^{\text{on-shell}}_{\text{eff}}$ represents only a local property of the on-shell distribution function around $\epsilon_{\textbf{k}}\simeq \mu_0$. The effective temperature $T_{\text{eff}}^{\text{on-shell}}$, from the numerical solution of Eq.~\eqref{F_QKE}, is shown in Fig.~\ref{fig:3}. For small $\nu_0$ and $g$, it coincides with the analytical expression $T_{\text{eff}}^{\text{on-shell}}=(1-\alpha)^{-1}T_{\text{cry}}$, with $\alpha=\frac{ g^2\nu_0\Delta T }{4\pi \gamma_0^2T_{\text{cry}}^2}$ derived from Eq.~(\ref{F_sol_approx}), as shown in the inset of~Fig.~\ref{fig:3}. Interestingly, from the numerical solution we observe a non-monotonic dependence of the $T^{\text{on-shell}}_{\text{eff}}$ on $\nu_0$. Whenever the physics of a phase transition is governed by the vicinity of the Fermi surface, we can thus predict  a non-monotonic behaviour of any critical temperature measured in terms of the cryostat temperature $T_{\text{cry}}$, as it is observed in the experiment \cite{1T-TaS2-Exp}.

\textit{Nonthermal Sommerfeld expansion}--- In a non-interacting Fermi gas at thermal equilibrium, the expectation values of a single-particle observable $ \hat{O}=\int dt \sum_{\textbf{k}}o_{\textbf{k}}(t) \psi^\dagger_{\textbf{k}}(t) \psi_{\textbf{k}}(t) $ at low temperature can be expressed as a power series of the temperature $T$ by means of the Sommerfeld expansion  \cite{Ashcroft76}. Due to the anti-symmetry property of the equilibrium distribution $F_0(\omega)$ around $\omega=\mu_0$, the leading order temperature-dependent term is always quadratic in temperature and linear in the energy derivative of the observable $(\pi^2 T^2/6)\partial_\epsilon [O(\epsilon)D(\epsilon)]_{\epsilon=\mu_0}$, where 
$O(\epsilon_\textbf{k})=o_{\textbf{k}}(\omega=\epsilon_\textbf{k})$
and $D(\epsilon)$ is the density of states. 
We will show now how this picture is modified in our two-bath setting, with the emergence of a leading-order  correction stemming from the nonthermal component of the fermionic distribution $F(\textbf{k},\omega)$. To describe this nonthermal contribution to time-independent single-particle observables, it is necessary to take into account the frequency dependence of the quasiparticle damping $\gamma(\omega)$ \footnote{For time-dependent observables, a constant electron damping $\gamma(\omega)=\gamma_0$ can generate a nonthermal correction, as shown in the Supplementary Material.}. In a setting with a finite electron damping, one also needs to define the frequency-integrated distribution $\tilde{F}(\epsilon)=\int \frac{d \omega}{\pi} A(\textbf{k},\omega) F(\epsilon,\omega)$, where $A(\textbf{k},\omega)$ is the spectral function of the electrons defined below Eq.~(\ref{QKE}). This $\tilde{F}(\epsilon)$ is the actual distribution that enters the Sommerfield expansion (in the standard case \cite{Ashcroft76} one typically has $\gamma \to 0$, which corresponds to $\tilde{F}(\epsilon)=F(\epsilon,\omega=\epsilon)$).
In the nonthermal case, $\tilde{F}(\epsilon)$ acquires a symmetric contribution around $\epsilon=\mu_0$, generating a new correction in the Sommerfeld expansion. To obtain a closed expression for this correction, we consider the usual starting point of the Sommerfeld expansion for a time-independent observable $\langle \hat{O}\rangle=\int \dd \epsilon \ O(\epsilon)\  D(\epsilon)\tilde{f}(\epsilon)$, where we defined $\tilde{f}=(1-\tilde{F})/2$.
We assume that the width $\gamma(\omega=\mu_0)$ of the spectral function is much smaller than the chemical potential $\mu_0$, and we further assume that $T_{\mathrm{cry}}\ll \mu_0$, as per the usual Sommerfeld expansion. In the limits of small $\nu_0$ and small $g$, we can use the analytical solution (\ref{F_sol_approx}) for the nonthermal distribution, and obtain the following  leading-order term quantifying the difference between the usual thermal expansion and the nonthermal case (see the Supplementary material for details of the derivation):
\begin{equation}
    \begin{split}
        \Delta \langle \hat{O} \rangle &\equiv \int \dd \epsilon \ O(\epsilon)\  D(\epsilon) [\tilde{f}(\epsilon)-\tilde{f}_0(\epsilon)] \\ &=\dfrac{g^2\nu_0 \Delta T}{4\pi\gamma_0^3} O(\mu_0)D(\mu_0)\partial_\omega \gamma(\omega)|_{\omega=\mu_0}.
    \end{split}
    \label{delta_O}
\end{equation}
We see that this new term is linear in the temperature difference $\Delta T = T_{\text{cav}}-T_{\text{cry}}$ and also proportional to the value of the observable evaluated at the chemical potential. It is also linearly proportional to the frequency-derivative of the electron damping.A term like \eqref{delta_O} is prohibited at thermal equilibrium due to the  anti-symmetry of $\tilde{F}(\epsilon)$ around $\epsilon=\mu_0$ excluding odd powers of $T_{\mathrm{cry}}$ from the expansion.

\textit{Conclusions}--- We have considered the nonthermal features of the electron distribution under the influence of two baths: one being the cryostat and the other the electromagnetic environment filtered by the presence of a Fabry-Perot cavity, as realized in state-of-the-art experimental setups. We found that the nonthermal electron distribution function possesses a reduced symmetry with respect the thermal distribution function. We have defined an effective temperature from the behaviour of the nonthermal on-shell distribution function near the Fermi surface, and shown that it depends non-monotonically on the cavity frequency, which is consistent with the non-monotonic behavior of critical temperature of the charge-density-wave transition observed in the experiment of Ref.~\cite{1T-TaS2-Exp}. In future work, we will  extend our investigation of the effects of the nonthermal electron distribution function to the critical temperature and gap equation. Moreover, we have shown that, in the Sommerfeld expansion of a single-particle observable, a nonthermal correction appears, which is proportional to the temperature difference between the two baths, to the observable evaluated at the Fermi surface, as well as to the frequency-derivative of the electron damping. The findings highlight the importance of taking into account nonthermal effects in two-bath settings through the modified fermionic distribution function, and how this generically can induce qualitative changes in  the material's properties.

\textit{Acknowledgements}--- We thank Martin Eckstein, Daniele Fausti, Denis Golez and Zala Lenarcic for fruitful discussions.
\bibliography{main.bib}

\begin{thebibliography}{34}%
\makeatletter
\providecommand \@ifxundefined [1]{%
 \@ifx{#1\undefined}
}%
\providecommand \@ifnum [1]{%
 \ifnum #1\expandafter \@firstoftwo
 \else \expandafter \@secondoftwo
 \fi
}%
\providecommand \@ifx [1]{%
 \ifx #1\expandafter \@firstoftwo
 \else \expandafter \@secondoftwo
 \fi
}%
\providecommand \natexlab [1]{#1}%
\providecommand \enquote  [1]{``#1''}%
\providecommand \bibnamefont  [1]{#1}%
\providecommand \bibfnamefont [1]{#1}%
\providecommand \citenamefont [1]{#1}%
\providecommand \href@noop [0]{\@secondoftwo}%
\providecommand \href [0]{\begingroup \@sanitize@url \@href}%
\providecommand \@href[1]{\@@startlink{#1}\@@href}%
\providecommand \@@href[1]{\endgroup#1\@@endlink}%
\providecommand \@sanitize@url [0]{\catcode `\\12\catcode `\$12\catcode
  `\&12\catcode `\#12\catcode `\^12\catcode `\_12\catcode `\%12\relax}%
\providecommand \@@startlink[1]{}%
\providecommand \@@endlink[0]{}%
\providecommand \url  [0]{\begingroup\@sanitize@url \@url }%
\providecommand \@url [1]{\endgroup\@href {#1}{\urlprefix }}%
\providecommand \urlprefix  [0]{URL }%
\providecommand \Eprint [0]{\href }%
\providecommand \doibase [0]{https://doi.org/}%
\providecommand \selectlanguage [0]{\@gobble}%
\providecommand \bibinfo  [0]{\@secondoftwo}%
\providecommand \bibfield  [0]{\@secondoftwo}%
\providecommand \translation [1]{[#1]}%
\providecommand \BibitemOpen [0]{}%
\providecommand \bibitemStop [0]{}%
\providecommand \bibitemNoStop [0]{.\EOS\space}%
\providecommand \EOS [0]{\spacefactor3000\relax}%
\providecommand \BibitemShut  [1]{\csname bibitem#1\endcsname}%
\let\auto@bib@innerbib\@empty
\bibitem [{\citenamefont {von Klitzing}\ \emph {et~al.}(2020)\citenamefont {von
  Klitzing}, \citenamefont {Chakraborty}, \citenamefont {Kim}, \citenamefont
  {Madhavan}, \citenamefont {Dai}, \citenamefont {McIver}, \citenamefont
  {Tokura}, \citenamefont {Savary}, \citenamefont {Smirnova}, \citenamefont
  {Rey}, \citenamefont {Felser}, \citenamefont {Gooth},\ and\ \citenamefont
  {Qi}}]{vonKlitzing2020}%
  \BibitemOpen
  \bibfield  {author} {\bibinfo {author} {\bibfnamefont {K.}~\bibnamefont {von
  Klitzing}}, \bibinfo {author} {\bibfnamefont {T.}~\bibnamefont
  {Chakraborty}}, \bibinfo {author} {\bibfnamefont {P.}~\bibnamefont {Kim}},
  \bibinfo {author} {\bibfnamefont {V.}~\bibnamefont {Madhavan}}, \bibinfo
  {author} {\bibfnamefont {X.}~\bibnamefont {Dai}}, \bibinfo {author}
  {\bibfnamefont {J.}~\bibnamefont {McIver}}, \bibinfo {author} {\bibfnamefont
  {Y.}~\bibnamefont {Tokura}}, \bibinfo {author} {\bibfnamefont
  {L.}~\bibnamefont {Savary}}, \bibinfo {author} {\bibfnamefont
  {D.}~\bibnamefont {Smirnova}}, \bibinfo {author} {\bibfnamefont {A.~M.}\
  \bibnamefont {Rey}}, \bibinfo {author} {\bibfnamefont {C.}~\bibnamefont
  {Felser}}, \bibinfo {author} {\bibfnamefont {J.}~\bibnamefont {Gooth}},\ and\
  \bibinfo {author} {\bibfnamefont {X.}~\bibnamefont {Qi}},\ }\bibfield
  {title} {\bibinfo {title} {40 years of the quantum hall effect},\ }\href
  {https://doi.org/10.1038/s42254-020-0209-1} {\bibfield  {journal} {\bibinfo
  {journal} {Nature Reviews Physics}\ }\textbf {\bibinfo {volume} {2}},\
  \bibinfo {pages} {397} (\bibinfo {year} {2020})}\BibitemShut {NoStop}%
\bibitem [{\citenamefont {Mackenbach}\ \emph {et~al.}(2022)\citenamefont
  {Mackenbach}, \citenamefont {Proll},\ and\ \citenamefont
  {Helander}}]{PhysRevLett.128.175001}%
  \BibitemOpen
  \bibfield  {author} {\bibinfo {author} {\bibfnamefont {R.~J.~J.}\
  \bibnamefont {Mackenbach}}, \bibinfo {author} {\bibfnamefont {J.~H.~E.}\
  \bibnamefont {Proll}},\ and\ \bibinfo {author} {\bibfnamefont
  {P.}~\bibnamefont {Helander}},\ }\bibfield  {title} {\bibinfo {title}
  {Available energy of trapped electrons and its relation to turbulent
  transport},\ }\href {https://doi.org/10.1103/PhysRevLett.128.175001}
  {\bibfield  {journal} {\bibinfo  {journal} {Phys. Rev. Lett.}\ }\textbf
  {\bibinfo {volume} {128}},\ \bibinfo {pages} {175001} (\bibinfo {year}
  {2022})}\BibitemShut {NoStop}%
\bibitem [{\citenamefont {Ren}\ \emph {et~al.}(2022)\citenamefont {Ren},
  \citenamefont {Lombez}, \citenamefont {Robert}, \citenamefont {Beret},
  \citenamefont {Lagarde}, \citenamefont {Urbaszek}, \citenamefont {Renucci},
  \citenamefont {Taniguchi}, \citenamefont {Watanabe}, \citenamefont
  {Crooker},\ and\ \citenamefont {Marie}}]{PhysRevLett.129.027402}%
  \BibitemOpen
  \bibfield  {author} {\bibinfo {author} {\bibfnamefont {L.}~\bibnamefont
  {Ren}}, \bibinfo {author} {\bibfnamefont {L.}~\bibnamefont {Lombez}},
  \bibinfo {author} {\bibfnamefont {C.}~\bibnamefont {Robert}}, \bibinfo
  {author} {\bibfnamefont {D.}~\bibnamefont {Beret}}, \bibinfo {author}
  {\bibfnamefont {D.}~\bibnamefont {Lagarde}}, \bibinfo {author} {\bibfnamefont
  {B.}~\bibnamefont {Urbaszek}}, \bibinfo {author} {\bibfnamefont
  {P.}~\bibnamefont {Renucci}}, \bibinfo {author} {\bibfnamefont
  {T.}~\bibnamefont {Taniguchi}}, \bibinfo {author} {\bibfnamefont
  {K.}~\bibnamefont {Watanabe}}, \bibinfo {author} {\bibfnamefont {S.~A.}\
  \bibnamefont {Crooker}},\ and\ \bibinfo {author} {\bibfnamefont
  {X.}~\bibnamefont {Marie}},\ }\bibfield  {title} {\bibinfo {title} {Optical
  detection of long electron spin transport lengths in a monolayer
  semiconductor},\ }\href {https://doi.org/10.1103/PhysRevLett.129.027402}
  {\bibfield  {journal} {\bibinfo  {journal} {Phys. Rev. Lett.}\ }\textbf
  {\bibinfo {volume} {129}},\ \bibinfo {pages} {027402} (\bibinfo {year}
  {2022})}\BibitemShut {NoStop}%
\bibitem [{\citenamefont {Ryu}\ and\ \citenamefont
  {Sim}(2022)}]{PhysRevLett.129.166801}%
  \BibitemOpen
  \bibfield  {author} {\bibinfo {author} {\bibfnamefont {S.}~\bibnamefont
  {Ryu}}\ and\ \bibinfo {author} {\bibfnamefont {H.-S.}\ \bibnamefont {Sim}},\
  }\bibfield  {title} {\bibinfo {title} {Partition of two interacting electrons
  by a potential barrier},\ }\href
  {https://doi.org/10.1103/PhysRevLett.129.166801} {\bibfield  {journal}
  {\bibinfo  {journal} {Phys. Rev. Lett.}\ }\textbf {\bibinfo {volume} {129}},\
  \bibinfo {pages} {166801} (\bibinfo {year} {2022})}\BibitemShut {NoStop}%
\bibitem [{\citenamefont {Needleman}\ and\ \citenamefont
  {Dogic}(2017)}]{Needleman2017}%
  \BibitemOpen
  \bibfield  {author} {\bibinfo {author} {\bibfnamefont {D.}~\bibnamefont
  {Needleman}}\ and\ \bibinfo {author} {\bibfnamefont {Z.}~\bibnamefont
  {Dogic}},\ }\bibfield  {title} {\bibinfo {title} {Active matter at the
  interface between materials science and cell biology},\ }\href
  {https://doi.org/10.1038/natrevmats.2017.48} {\bibfield  {journal} {\bibinfo
  {journal} {Nature Reviews Materials}\ }\textbf {\bibinfo {volume} {2}},\
  \bibinfo {pages} {17048} (\bibinfo {year} {2017})}\BibitemShut {NoStop}%
\bibitem [{\citenamefont {Popkin}(2016)}]{Popkin2016}%
  \BibitemOpen
  \bibfield  {author} {\bibinfo {author} {\bibfnamefont {G.}~\bibnamefont
  {Popkin}},\ }\bibfield  {title} {\bibinfo {title} {The physics of life},\
  }\href {https://doi.org/10.1038/529016a} {\bibfield  {journal} {\bibinfo
  {journal} {Nature}\ }\textbf {\bibinfo {volume} {529}},\ \bibinfo {pages}
  {16} (\bibinfo {year} {2016})}\BibitemShut {NoStop}%
\bibitem [{\citenamefont {Xi}\ \emph {et~al.}(2019)\citenamefont {Xi},
  \citenamefont {Saw}, \citenamefont {Delacour}, \citenamefont {Lim},\ and\
  \citenamefont {Ladoux}}]{Xi2019}%
  \BibitemOpen
  \bibfield  {author} {\bibinfo {author} {\bibfnamefont {W.}~\bibnamefont
  {Xi}}, \bibinfo {author} {\bibfnamefont {T.~B.}\ \bibnamefont {Saw}},
  \bibinfo {author} {\bibfnamefont {D.}~\bibnamefont {Delacour}}, \bibinfo
  {author} {\bibfnamefont {C.~T.}\ \bibnamefont {Lim}},\ and\ \bibinfo {author}
  {\bibfnamefont {B.}~\bibnamefont {Ladoux}},\ }\bibfield  {title} {\bibinfo
  {title} {Material approaches to active tissue mechanics},\ }\href
  {https://doi.org/10.1038/s41578-018-0066-z} {\bibfield  {journal} {\bibinfo
  {journal} {Nature Reviews Materials}\ }\textbf {\bibinfo {volume} {4}},\
  \bibinfo {pages} {23} (\bibinfo {year} {2019})}\BibitemShut {NoStop}%
\bibitem [{\citenamefont {Foglino}\ \emph {et~al.}(2019)\citenamefont
  {Foglino}, \citenamefont {Locatelli}, \citenamefont {Brackley}, \citenamefont
  {Michieletto}, \citenamefont {Likos},\ and\ \citenamefont
  {Marenduzzo}}]{AM-nonequ}%
  \BibitemOpen
  \bibfield  {author} {\bibinfo {author} {\bibfnamefont {M.}~\bibnamefont
  {Foglino}}, \bibinfo {author} {\bibfnamefont {E.}~\bibnamefont {Locatelli}},
  \bibinfo {author} {\bibfnamefont {C.~A.}\ \bibnamefont {Brackley}}, \bibinfo
  {author} {\bibfnamefont {D.}~\bibnamefont {Michieletto}}, \bibinfo {author}
  {\bibfnamefont {C.~N.}\ \bibnamefont {Likos}},\ and\ \bibinfo {author}
  {\bibfnamefont {D.}~\bibnamefont {Marenduzzo}},\ }\bibfield  {title}
  {\bibinfo {title} {Non-equilibrium effects of molecular motors on polymers},\
  }\href {https://doi.org/10.1039/C9SM00273A} {\bibfield  {journal} {\bibinfo
  {journal} {Soft Matter}\ }\textbf {\bibinfo {volume} {15}},\ \bibinfo {pages}
  {5995} (\bibinfo {year} {2019})}\BibitemShut {NoStop}%
\bibitem [{\citenamefont {Pacchioni}(2016)}]{Pacchioni2016}%
  \BibitemOpen
  \bibfield  {author} {\bibinfo {author} {\bibfnamefont {G.}~\bibnamefont
  {Pacchioni}},\ }\bibfield  {title} {\bibinfo {title} {Molecular motors: You
  spin me round},\ }\href {https://doi.org/10.1038/natrevmats.2016.45}
  {\bibfield  {journal} {\bibinfo  {journal} {Nature Reviews Materials}\
  }\textbf {\bibinfo {volume} {1}},\ \bibinfo {pages} {16045} (\bibinfo {year}
  {2016})}\BibitemShut {NoStop}%
\bibitem [{\citenamefont {Pomeau}(2016)}]{Pomeau2016}%
  \BibitemOpen
  \bibfield  {author} {\bibinfo {author} {\bibfnamefont {Y.}~\bibnamefont
  {Pomeau}},\ }\bibfield  {title} {\bibinfo {title} {The long and winding
  road},\ }\href {https://doi.org/10.1038/nphys3684} {\bibfield  {journal}
  {\bibinfo  {journal} {Nature Physics}\ }\textbf {\bibinfo {volume} {12}},\
  \bibinfo {pages} {198} (\bibinfo {year} {2016})}\BibitemShut {NoStop}%
\bibitem [{\citenamefont {Lemoult}\ \emph {et~al.}(2016)\citenamefont
  {Lemoult}, \citenamefont {Shi}, \citenamefont {Avila}, \citenamefont
  {Jalikop}, \citenamefont {Avila},\ and\ \citenamefont {Hof}}]{Lemoult2016}%
  \BibitemOpen
  \bibfield  {author} {\bibinfo {author} {\bibfnamefont {G.}~\bibnamefont
  {Lemoult}}, \bibinfo {author} {\bibfnamefont {L.}~\bibnamefont {Shi}},
  \bibinfo {author} {\bibfnamefont {K.}~\bibnamefont {Avila}}, \bibinfo
  {author} {\bibfnamefont {S.~V.}\ \bibnamefont {Jalikop}}, \bibinfo {author}
  {\bibfnamefont {M.}~\bibnamefont {Avila}},\ and\ \bibinfo {author}
  {\bibfnamefont {B.}~\bibnamefont {Hof}},\ }\bibfield  {title} {\bibinfo
  {title} {Directed percolation phase transition to sustained turbulence in
  couette flow},\ }\href {https://doi.org/10.1038/nphys3675} {\bibfield
  {journal} {\bibinfo  {journal} {Nature Physics}\ }\textbf {\bibinfo {volume}
  {12}},\ \bibinfo {pages} {254} (\bibinfo {year} {2016})}\BibitemShut
  {NoStop}%
\bibitem [{\citenamefont {Oberlack}\ \emph {et~al.}(2022)\citenamefont
  {Oberlack}, \citenamefont {Hoyas}, \citenamefont {Kraheberger}, \citenamefont
  {Alc\'antara-\'Avila},\ and\ \citenamefont {Laux}}]{PhysRevLett.128.024502}%
  \BibitemOpen
  \bibfield  {author} {\bibinfo {author} {\bibfnamefont {M.}~\bibnamefont
  {Oberlack}}, \bibinfo {author} {\bibfnamefont {S.}~\bibnamefont {Hoyas}},
  \bibinfo {author} {\bibfnamefont {S.~V.}\ \bibnamefont {Kraheberger}},
  \bibinfo {author} {\bibfnamefont {F.}~\bibnamefont {Alc\'antara-\'Avila}},\
  and\ \bibinfo {author} {\bibfnamefont {J.}~\bibnamefont {Laux}},\ }\bibfield
  {title} {\bibinfo {title} {Turbulence statistics of arbitrary moments of
  wall-bounded shear flows: A symmetry approach},\ }\href
  {https://doi.org/10.1103/PhysRevLett.128.024502} {\bibfield  {journal}
  {\bibinfo  {journal} {Phys. Rev. Lett.}\ }\textbf {\bibinfo {volume} {128}},\
  \bibinfo {pages} {024502} (\bibinfo {year} {2022})}\BibitemShut {NoStop}%
\bibitem [{\citenamefont {Siegman}(1986)}]{siegman1986lasers}%
  \BibitemOpen
  \bibfield  {author} {\bibinfo {author} {\bibfnamefont {A.~E.}\ \bibnamefont
  {Siegman}},\ }\href@noop {} {\emph {\bibinfo {title} {Lasers}}}\ (\bibinfo
  {publisher} {University science books},\ \bibinfo {year} {1986})\BibitemShut
  {NoStop}%
\bibitem [{\citenamefont {Garcia-Vidal}\ \emph {et~al.}(2021)\citenamefont
  {Garcia-Vidal}, \citenamefont {Ciuti},\ and\ \citenamefont
  {Ebbesen}}]{garcia2021manipulating}%
  \BibitemOpen
  \bibfield  {author} {\bibinfo {author} {\bibfnamefont {F.~J.}\ \bibnamefont
  {Garcia-Vidal}}, \bibinfo {author} {\bibfnamefont {C.}~\bibnamefont
  {Ciuti}},\ and\ \bibinfo {author} {\bibfnamefont {T.~W.}\ \bibnamefont
  {Ebbesen}},\ }\bibfield  {title} {\bibinfo {title} {Manipulating matter by
  strong coupling to vacuum fields},\ }\href
  {https://doi.org/10.1126/science.abd0336} {\bibfield  {journal} {\bibinfo
  {journal} {Science}\ }\textbf {\bibinfo {volume} {373}},\  (\bibinfo {year}
  {2021})}\BibitemShut {NoStop}%
\bibitem [{\citenamefont {Mivehvar}\ \emph {et~al.}(2021)\citenamefont
  {Mivehvar}, \citenamefont {Piazza}, \citenamefont {Donner},\ and\
  \citenamefont {Ritsch}}]{mivehvar2021cavity}%
  \BibitemOpen
  \bibfield  {author} {\bibinfo {author} {\bibfnamefont {F.}~\bibnamefont
  {Mivehvar}}, \bibinfo {author} {\bibfnamefont {F.}~\bibnamefont {Piazza}},
  \bibinfo {author} {\bibfnamefont {T.}~\bibnamefont {Donner}},\ and\ \bibinfo
  {author} {\bibfnamefont {H.}~\bibnamefont {Ritsch}},\ }\bibfield  {title}
  {\bibinfo {title} {Cavity qed with quantum gases: new paradigms in many-body
  physics},\ }\href {https://doi.org/10.1080/00018732.2021.1969727} {\bibfield
  {journal} {\bibinfo  {journal} {Advances in Physics}\ }\textbf {\bibinfo
  {volume} {70}},\ \bibinfo {pages} {1} (\bibinfo {year} {2021})}\BibitemShut
  {NoStop}%
\bibitem [{\citenamefont {Schlawin}\ \emph {et~al.}(2022)\citenamefont
  {Schlawin}, \citenamefont {Kennes},\ and\ \citenamefont
  {Sentef}}]{schlawin2022cavity}%
  \BibitemOpen
  \bibfield  {author} {\bibinfo {author} {\bibfnamefont {F.}~\bibnamefont
  {Schlawin}}, \bibinfo {author} {\bibfnamefont {D.~M.}\ \bibnamefont
  {Kennes}},\ and\ \bibinfo {author} {\bibfnamefont {M.~A.}\ \bibnamefont
  {Sentef}},\ }\bibfield  {title} {\bibinfo {title} {Cavity quantum
  materials},\ }\href {https://doi.org/10.1063/5.0083825} {\bibfield  {journal}
  {\bibinfo  {journal} {Applied Physics Reviews}\ }\textbf {\bibinfo {volume}
  {9}},\  (\bibinfo {year} {2022})}\BibitemShut {NoStop}%
\bibitem [{\citenamefont {Bloch}\ \emph {et~al.}(2022)\citenamefont {Bloch},
  \citenamefont {Cavalleri}, \citenamefont {Galitski}, \citenamefont {Hafezi},\
  and\ \citenamefont {Rubio}}]{bloch2022strongly}%
  \BibitemOpen
  \bibfield  {author} {\bibinfo {author} {\bibfnamefont {J.}~\bibnamefont
  {Bloch}}, \bibinfo {author} {\bibfnamefont {A.}~\bibnamefont {Cavalleri}},
  \bibinfo {author} {\bibfnamefont {V.}~\bibnamefont {Galitski}}, \bibinfo
  {author} {\bibfnamefont {M.}~\bibnamefont {Hafezi}},\ and\ \bibinfo {author}
  {\bibfnamefont {A.}~\bibnamefont {Rubio}},\ }\bibfield  {title} {\bibinfo
  {title} {Strongly correlated electron--photon systems},\ }\href
  {https://doi.org/10.1038/s41586-022-04726-w} {\bibfield  {journal} {\bibinfo
  {journal} {Nature}\ }\textbf {\bibinfo {volume} {606}},\ \bibinfo {pages}
  {41} (\bibinfo {year} {2022})}\BibitemShut {NoStop}%
\bibitem [{\citenamefont {De~La~Torre}\ \emph {et~al.}(2021)\citenamefont
  {De~La~Torre}, \citenamefont {Kennes}, \citenamefont {Claassen},
  \citenamefont {Gerber}, \citenamefont {McIver},\ and\ \citenamefont
  {Sentef}}]{de2021colloquium}%
  \BibitemOpen
  \bibfield  {author} {\bibinfo {author} {\bibfnamefont {A.}~\bibnamefont
  {De~La~Torre}}, \bibinfo {author} {\bibfnamefont {D.~M.}\ \bibnamefont
  {Kennes}}, \bibinfo {author} {\bibfnamefont {M.}~\bibnamefont {Claassen}},
  \bibinfo {author} {\bibfnamefont {S.}~\bibnamefont {Gerber}}, \bibinfo
  {author} {\bibfnamefont {J.~W.}\ \bibnamefont {McIver}},\ and\ \bibinfo
  {author} {\bibfnamefont {M.~A.}\ \bibnamefont {Sentef}},\ }\bibfield  {title}
  {\bibinfo {title} {Colloquium: Nonthermal pathways to ultrafast control in
  quantum materials},\ }\href {https://doi.org/10.1103/RevModPhys.93.041002}
  {\bibfield  {journal} {\bibinfo  {journal} {Reviews of Modern Physics}\
  }\textbf {\bibinfo {volume} {93}},\ \bibinfo {pages} {041002} (\bibinfo
  {year} {2021})}\BibitemShut {NoStop}%
\bibitem [{\citenamefont {Curtis}\ \emph {et~al.}(2019)\citenamefont {Curtis},
  \citenamefont {Raines}, \citenamefont {Allocca}, \citenamefont {Hafezi},\
  and\ \citenamefont {Galitski}}]{curtis2019cavity}%
  \BibitemOpen
  \bibfield  {author} {\bibinfo {author} {\bibfnamefont {J.~B.}\ \bibnamefont
  {Curtis}}, \bibinfo {author} {\bibfnamefont {Z.~M.}\ \bibnamefont {Raines}},
  \bibinfo {author} {\bibfnamefont {A.~A.}\ \bibnamefont {Allocca}}, \bibinfo
  {author} {\bibfnamefont {M.}~\bibnamefont {Hafezi}},\ and\ \bibinfo {author}
  {\bibfnamefont {V.~M.}\ \bibnamefont {Galitski}},\ }\bibfield  {title}
  {\bibinfo {title} {Cavity quantum eliashberg enhancement of
  superconductivity},\ }\href {https://doi.org/10.1103/PhysRevLett.122.167002}
  {\bibfield  {journal} {\bibinfo  {journal} {Physical review letters}\
  }\textbf {\bibinfo {volume} {122}},\ \bibinfo {pages} {167002} (\bibinfo
  {year} {2019})}\BibitemShut {NoStop}%
\bibitem [{\citenamefont {Chakraborty}\ and\ \citenamefont
  {Piazza}(2021)}]{PhysRevLett.127.177002}%
  \BibitemOpen
  \bibfield  {author} {\bibinfo {author} {\bibfnamefont {A.}~\bibnamefont
  {Chakraborty}}\ and\ \bibinfo {author} {\bibfnamefont {F.}~\bibnamefont
  {Piazza}},\ }\bibfield  {title} {\bibinfo {title} {Long-range photon
  fluctuations enhance photon-mediated electron pairing and
  superconductivity},\ }\href {https://doi.org/10.1103/PhysRevLett.127.177002}
  {\bibfield  {journal} {\bibinfo  {journal} {Phys. Rev. Lett.}\ }\textbf
  {\bibinfo {volume} {127}},\ \bibinfo {pages} {177002} (\bibinfo {year}
  {2021})}\BibitemShut {NoStop}%
\bibitem [{\citenamefont {Chakraborty}\ and\ \citenamefont
  {Piazza}(2022)}]{chakraborty2022controlling}%
  \BibitemOpen
  \bibfield  {author} {\bibinfo {author} {\bibfnamefont {A.}~\bibnamefont
  {Chakraborty}}\ and\ \bibinfo {author} {\bibfnamefont {F.}~\bibnamefont
  {Piazza}},\ }\bibfield  {title} {\bibinfo {title} {Controlling collective
  phenomena by engineering the quantum state of force carriers: The case of
  photon-mediated superconductivity and its criticality},\ }\bibfield
  {journal} {\bibinfo  {journal} {arXiv preprint}\ }\href
  {https://doi.org/10.48550/arXiv.2207.07131} {10.48550/arXiv.2207.07131}
  (\bibinfo {year} {2022})\BibitemShut {NoStop}%
\bibitem [{\citenamefont {Eckhardt}\ \emph {et~al.}(2024)\citenamefont
  {Eckhardt}, \citenamefont {Chattopadhyay}, \citenamefont {Kennes},
  \citenamefont {Demler}, \citenamefont {Sentef},\ and\ \citenamefont
  {Michael}}]{eckhardt2024theory}%
  \BibitemOpen
  \bibfield  {author} {\bibinfo {author} {\bibfnamefont {C.~J.}\ \bibnamefont
  {Eckhardt}}, \bibinfo {author} {\bibfnamefont {S.}~\bibnamefont
  {Chattopadhyay}}, \bibinfo {author} {\bibfnamefont {D.~M.}\ \bibnamefont
  {Kennes}}, \bibinfo {author} {\bibfnamefont {E.~A.}\ \bibnamefont {Demler}},
  \bibinfo {author} {\bibfnamefont {M.~A.}\ \bibnamefont {Sentef}},\ and\
  \bibinfo {author} {\bibfnamefont {M.~H.}\ \bibnamefont {Michael}},\
  }\bibfield  {title} {\bibinfo {title} {Theory of resonantly enhanced
  photo-induced superconductivity},\ }\href
  {https://doi.org/10.1038/s41467-024-46632-x} {\bibfield  {journal} {\bibinfo
  {journal} {Nature Communications}\ }\textbf {\bibinfo {volume} {15}},\
  \bibinfo {pages} {2300} (\bibinfo {year} {2024})}\BibitemShut {NoStop}%
\bibitem [{\citenamefont {Vi{\~n}as~Bostr{\"o}m}\ \emph
  {et~al.}(2023)\citenamefont {Vi{\~n}as~Bostr{\"o}m}, \citenamefont {Sriram},
  \citenamefont {Claassen},\ and\ \citenamefont
  {Rubio}}]{vinas2023controlling}%
  \BibitemOpen
  \bibfield  {author} {\bibinfo {author} {\bibfnamefont {E.}~\bibnamefont
  {Vi{\~n}as~Bostr{\"o}m}}, \bibinfo {author} {\bibfnamefont {A.}~\bibnamefont
  {Sriram}}, \bibinfo {author} {\bibfnamefont {M.}~\bibnamefont {Claassen}},\
  and\ \bibinfo {author} {\bibfnamefont {A.}~\bibnamefont {Rubio}},\ }\bibfield
   {title} {\bibinfo {title} {Controlling the magnetic state of the proximate
  quantum spin liquid $\alpha$-rucl3 with an optical cavity},\ }\href
  {https://doi.org/10.1038/s41524-023-01158-6} {\bibfield  {journal} {\bibinfo
  {journal} {npj Computational Materials}\ }\textbf {\bibinfo {volume} {9}},\
  \bibinfo {pages} {202} (\bibinfo {year} {2023})}\BibitemShut {NoStop}%
\bibitem [{\citenamefont {Jarc}\ \emph {et~al.}(2023)\citenamefont {Jarc},
  \citenamefont {Mathengattil}, \citenamefont {Montanaro}, \citenamefont
  {Giusti}, \citenamefont {Rigoni}, \citenamefont {Sergo}, \citenamefont
  {Fassioli}, \citenamefont {Winnerl}, \citenamefont {Dal~Zilio}, \citenamefont
  {Mihailovic} \emph {et~al.}}]{1T-TaS2-Exp}%
  \BibitemOpen
  \bibfield  {author} {\bibinfo {author} {\bibfnamefont {G.}~\bibnamefont
  {Jarc}}, \bibinfo {author} {\bibfnamefont {S.~Y.}\ \bibnamefont
  {Mathengattil}}, \bibinfo {author} {\bibfnamefont {A.}~\bibnamefont
  {Montanaro}}, \bibinfo {author} {\bibfnamefont {F.}~\bibnamefont {Giusti}},
  \bibinfo {author} {\bibfnamefont {E.~M.}\ \bibnamefont {Rigoni}}, \bibinfo
  {author} {\bibfnamefont {R.}~\bibnamefont {Sergo}}, \bibinfo {author}
  {\bibfnamefont {F.}~\bibnamefont {Fassioli}}, \bibinfo {author}
  {\bibfnamefont {S.}~\bibnamefont {Winnerl}}, \bibinfo {author} {\bibfnamefont
  {S.}~\bibnamefont {Dal~Zilio}}, \bibinfo {author} {\bibfnamefont
  {D.}~\bibnamefont {Mihailovic}}, \emph {et~al.},\ }\bibfield  {title}
  {\bibinfo {title} {Cavity-mediated thermal control of metal-to-insulator
  transition in $\text{1T}-\text{Ta} \text{S}_2$},\ }\href
  {https://doi.org/10.1038/s41586-023-06596-2} {\bibfield  {journal} {\bibinfo
  {journal} {Nature}\ }\textbf {\bibinfo {volume} {622}},\ \bibinfo {pages}
  {487} (\bibinfo {year} {2023})}\BibitemShut {NoStop}%
\bibitem [{\citenamefont {Chiriac{\`o}}(2023)}]{chiriaco2023thermal}%
  \BibitemOpen
  \bibfield  {author} {\bibinfo {author} {\bibfnamefont {G.}~\bibnamefont
  {Chiriac{\`o}}},\ }\bibfield  {title} {\bibinfo {title} {Thermal purcell
  effect and cavity-induced renormalization of dissipations},\ }\bibfield
  {journal} {\bibinfo  {journal} {arXiv preprint}\ }\href
  {https://doi.org/10.48550/arXiv.2310.15184} {10.48550/arXiv.2310.15184}
  (\bibinfo {year} {2023})\BibitemShut {NoStop}%
\bibitem [{\citenamefont {Eliashberg}(1970)}]{eliashberg1970film}%
  \BibitemOpen
  \bibfield  {author} {\bibinfo {author} {\bibfnamefont {G.}~\bibnamefont
  {Eliashberg}},\ }\href@noop {} {\emph {\bibinfo {title} {Film
  superconductivity stimulated by a high-frequency field.}}},\ \bibinfo {type}
  {Tech. Rep.}\ (\bibinfo  {institution} {Inst. of Theoretical Physics,
  Moscow},\ \bibinfo {year} {1970})\BibitemShut {NoStop}%
\bibitem [{\citenamefont {Kamenev}(2023)}]{kamenev2023field}%
  \BibitemOpen
  \bibfield  {author} {\bibinfo {author} {\bibfnamefont {A.}~\bibnamefont
  {Kamenev}},\ }\href@noop {} {\emph {\bibinfo {title} {Field theory of
  non-equilibrium systems}}}\ (\bibinfo  {publisher} {Cambridge University
  Press},\ \bibinfo {year} {2023})\BibitemShut {NoStop}%
\bibitem [{\citenamefont {Sieberer}\ \emph {et~al.}(2016)\citenamefont
  {Sieberer}, \citenamefont {Buchhold},\ and\ \citenamefont
  {Diehl}}]{Keldysh-OQS}%
  \BibitemOpen
  \bibfield  {author} {\bibinfo {author} {\bibfnamefont {L.~M.}\ \bibnamefont
  {Sieberer}}, \bibinfo {author} {\bibfnamefont {M.}~\bibnamefont {Buchhold}},\
  and\ \bibinfo {author} {\bibfnamefont {S.}~\bibnamefont {Diehl}},\ }\bibfield
   {title} {\bibinfo {title} {Keldysh field theory for driven open quantum
  systems},\ }\href {https://doi.org/10.1088/0034-4885/79/9/096001} {\bibfield
  {journal} {\bibinfo  {journal} {Reports on Progress in Physics}\ }\textbf
  {\bibinfo {volume} {79}},\ \bibinfo {pages} {096001} (\bibinfo {year}
  {2016})}\BibitemShut {NoStop}%
\bibitem [{\citenamefont {Schlawin}\ \emph {et~al.}(2019)\citenamefont
  {Schlawin}, \citenamefont {Cavalleri},\ and\ \citenamefont
  {Jaksch}}]{Cavalleri-SC}%
  \BibitemOpen
  \bibfield  {author} {\bibinfo {author} {\bibfnamefont {F.}~\bibnamefont
  {Schlawin}}, \bibinfo {author} {\bibfnamefont {A.}~\bibnamefont
  {Cavalleri}},\ and\ \bibinfo {author} {\bibfnamefont {D.}~\bibnamefont
  {Jaksch}},\ }\bibfield  {title} {\bibinfo {title} {Cavity-mediated
  electron-photon superconductivity},\ }\href
  {https://doi.org/10.1103/PhysRevLett.122.133602} {\bibfield  {journal}
  {\bibinfo  {journal} {Phys. Rev. Lett.}\ }\textbf {\bibinfo {volume} {122}},\
  \bibinfo {pages} {133602} (\bibinfo {year} {2019})}\BibitemShut {NoStop}%
\bibitem [{\citenamefont {Andolina}\ \emph {et~al.}(2024)\citenamefont
  {Andolina}, \citenamefont {De~Pasquale}, \citenamefont {Pellegrino},
  \citenamefont {Torre}, \citenamefont {Koppens},\ and\ \citenamefont
  {Polini}}]{andolina2024can}%
  \BibitemOpen
  \bibfield  {author} {\bibinfo {author} {\bibfnamefont {G.~M.}\ \bibnamefont
  {Andolina}}, \bibinfo {author} {\bibfnamefont {A.}~\bibnamefont
  {De~Pasquale}}, \bibinfo {author} {\bibfnamefont {F.~M.~D.}\ \bibnamefont
  {Pellegrino}}, \bibinfo {author} {\bibfnamefont {I.}~\bibnamefont {Torre}},
  \bibinfo {author} {\bibfnamefont {F.~H.~L.}\ \bibnamefont {Koppens}},\ and\
  \bibinfo {author} {\bibfnamefont {M.}~\bibnamefont {Polini}},\ }\bibfield
  {title} {\bibinfo {title} {Amperean superconductivity cannot be induced by
  deep subwavelength cavities in a two-dimensional material},\ }\href
  {https://doi.org/10.1103/PhysRevB.109.104513} {\bibfield  {journal} {\bibinfo
   {journal} {Phys. Rev. B}\ }\textbf {\bibinfo {volume} {109}},\ \bibinfo
  {pages} {104513} (\bibinfo {year} {2024})}\BibitemShut {NoStop}%
\bibitem [{\citenamefont {Piazza}\ and\ \citenamefont
  {Strack}(2014)}]{piazza2014quantum}%
  \BibitemOpen
  \bibfield  {author} {\bibinfo {author} {\bibfnamefont {F.}~\bibnamefont
  {Piazza}}\ and\ \bibinfo {author} {\bibfnamefont {P.}~\bibnamefont
  {Strack}},\ }\bibfield  {title} {\bibinfo {title} {Quantum kinetics of
  ultracold fermions coupled to an optical resonator},\ }\href
  {https://doi.org/10.1103/PhysRevA.90.043823} {\bibfield  {journal} {\bibinfo
  {journal} {Physical Review A}\ }\textbf {\bibinfo {volume} {90}},\ \bibinfo
  {pages} {043823} (\bibinfo {year} {2014})}\BibitemShut {NoStop}%
\bibitem [{\citenamefont {Rao}\ and\ \citenamefont
  {Piazza}(2023)}]{Rao-Francesco}%
  \BibitemOpen
  \bibfield  {author} {\bibinfo {author} {\bibfnamefont {P.}~\bibnamefont
  {Rao}}\ and\ \bibinfo {author} {\bibfnamefont {F.}~\bibnamefont {Piazza}},\
  }\bibfield  {title} {\bibinfo {title} {Non-fermi-liquid behavior from cavity
  electromagnetic vacuum fluctuations at the superradiant transition},\ }\href
  {https://doi.org/10.1103/PhysRevLett.130.083603} {\bibfield  {journal}
  {\bibinfo  {journal} {Phys. Rev. Lett.}\ }\textbf {\bibinfo {volume} {130}},\
  \bibinfo {pages} {083603} (\bibinfo {year} {2023})}\BibitemShut {NoStop}%
\bibitem [{\citenamefont {Ashcroft}\ and\ \citenamefont
  {Mermin}(1976)}]{Ashcroft76}%
  \BibitemOpen
  \bibfield  {author} {\bibinfo {author} {\bibfnamefont {N.~W.}\ \bibnamefont
  {Ashcroft}}\ and\ \bibinfo {author} {\bibfnamefont {N.~D.}\ \bibnamefont
  {Mermin}},\ }\href@noop {} {\emph {\bibinfo {title} {{S}olid {S}tate
  {P}hysics}}}\ (\bibinfo  {publisher} {Holt-Saunders},\ \bibinfo {year}
  {1976})\BibitemShut {NoStop}%
\bibitem [{Note1()}]{Note1}%
  \BibitemOpen
  \bibinfo {note} {For time-dependent observables, a constant electron damping
  $\gamma (\omega )=\gamma _0$ can generate a nonthermal correction, as shown
  in the Supplementary Material.}\BibitemShut {Stop}%
\end{thebibliography}%

\makeatletter

\renewcommand{\t}[1]{\text{#1}}
\renewcommand{\theequation}{S\arabic{equation}}
\renewcommand{\selectlanguage}[1]{}
\renewcommand{\thefigure}{S\arabic{figure}}

\setcounter{equation}{0}
\setcounter{secnumdepth}{1}

\renewcommand{\thesection}{S\arabic{section}}
\onecolumngrid

\newpage

\begin{center}
  \textbf{\large Supplementary material for ``Nonthermal electron-photon steady states in open cavity quantum materials''}\\[.2cm]
  R. Flores-Calderón,$^{1,*}$, Md Mursalin Islam $^{1,*}$, Michele Pini$^{1,*}$, Francesco Piazza,$^{2,1}$ \\[.1cm]
  {\itshape
  ${}^1$ Max Planck Institute for the Physics of Complex Systems, Nöthnitzer Stra{\ss}e 38, 01187 Dresden, Germany\\
  ${}^2$Theoretical Physics III, Center for Electronic Correlations and Magnetism,
Institute of Physics, University of Augsburg, 86135 Augsburg, Germany\\}
  ${}^*$Electronic address: rflorescalderon@pks.mpg.de\\
(Dated: \today)\\[1cm]
\end{center}

\section{ Photon field coupled to a thermal bath}
Starting with the general framework of a real photon field, which has $a$ as an annihilation operator, coupled to a bath we may model the bath as multiple real photon fields with different dispersions. The Keldish action for the photon field will be given generically by:

\begin{align}
    S_\phi=&\sum_{\textbf{k}}  \oint_{\mathcal{C}} L dt \ \abs{\dot{\phi}(\textbf{k},t)}^2-\nu(\textbf{k})^2\abs{\phi(\textbf{k},t)}^2=\sum_{\textbf{k}}L \int_{-\infty}^{\infty}dt \ \abs{\dot{\phi}_+(\textbf{k},t)}^2-\nu(\textbf{k})^2\abs{\phi_+(\textbf{k},t)}^2 -\abs{\dot{\phi}_-(\textbf{k},t)}^2+\nu(\textbf{k})^2\abs{\phi_-(\textbf{k},t)}^2\\
    &=\sum_{\textbf{k}}\int_{-\infty}^{\infty} d\omega L \begin{pmatrix}
    \phi^{cl}(\textbf{k},\omega)^* && \phi^{q}(\textbf{k},\omega)^*
    \end{pmatrix}
    \begin{pmatrix}
    0 &&(\omega-i0)^2-\nu(\textbf{k})^2\\
    (\omega+i0)^2-\nu(\textbf{k})^2 && 2i0F(\omega)
    \end{pmatrix}
    \begin{pmatrix}
    \phi^{cl}(\textbf{k},\omega) \\ \phi^{q}(\textbf{k},\omega)
    \end{pmatrix},
\end{align}
 where the first line describes the field along the Keldysh contour being expressed in terms of the $+,-$ branches. We have also introduced a length which has units of inverse energy and in cavities comes from the $z$ axis boundary conditions that quantize the energy, this also has the added benefit of giving the field $\phi(x,t)$ dimensionless units,we have the Fourier convention $\phi(t)=\int \frac{\dd \omega}{\sqrt{2\pi}}\phi(\omega)e^{-i\omega t}$. The second line is fourier transformed for each branch and a Keldysh rotation has been applied meaning $\phi^{cl,q}=(\phi_+\pm\phi_-)/2$ and the infinitesimal $\pm i 0$ are needed to express the physical coupling of the two branches at $t=-\infty$ given by the initial density matrix. $F(\omega)$ is likewise a distribution function for the correct normalization which will be replaced by the finite coupling to the bath. The bath will have the same structure but for multiple frequencies meaning an action of the form:
\begin{align}
    S_\theta=\sum_{\textbf{k},s}\int_{-\infty}^{\infty} d\omega L \begin{pmatrix}
    \theta_s^{cl}(\textbf{k},\omega)^* && \theta_s^{q}(\textbf{k},\omega)^*
    \end{pmatrix}
    \begin{pmatrix}
    0 &&(\omega-i0)^2-\nu_s(\textbf{k})^2\\
    (\omega+i0)^2-\nu_s(\textbf{k})^2 && 2i0D(\omega)
    \end{pmatrix}
    \begin{pmatrix}
    \theta_s^{cl}(\textbf{k},\omega) \\ \theta_s^{q}(\textbf{k},\omega)
    \end{pmatrix},\label{freephotonprop}
\end{align}
 where $\theta_s^{q}(\textbf{k},\omega)$ are now the bath variables and $B_0(\omega)=\coth(\omega/2T_{\text{cav}})$ is the distribution function assuming the bath is in thermal equilibrium, which not necessarily implies the $\phi$ field is. To model the coupling between the bath and field we take the simple interaction:
\begin{align}
    S_{\theta\phi}=\sum_{\textbf{k},s}\oint_{\mathcal{C}}dt \ L t_s^{ph} \phi(\textbf{k},t) \theta_s(\textbf{k},t)=\sum_{\textbf{k},s}\int_{-\infty}^{\infty} L d\omega \ t_s^{ph} \Phi(\textbf{k},\omega)^\dagger\tau_x\Theta_s(\textbf{k},\omega),
\end{align}
where the coupling changes for each field and is dimensionless by virtue of the the $L$ factor the final vector fields are in Keldysh space meaning $\Phi(\textbf{k},\omega)^T=\begin{pmatrix}
    \phi^{cl}(\textbf{k},\omega) && \phi^{q}(\textbf{k},\omega)
    \end{pmatrix}$ and $\Theta_s(\textbf{k},\omega)= \begin{pmatrix}
    \theta_s^{cl}(\textbf{k},\omega) && \theta_s^{q}(\textbf{k},\omega)
    \end{pmatrix}$. The matrix $\tau_x$ is just the first Pauli matrix in the Keldysh space. Now if one wants to obtain properties of the stationary state but arbitrary long time correlations then the Keldysh partition function is expressed as:
    \begin{align}
        Z_\phi= \dfrac{\Tr{\rho(t_f\rightarrow\infty)}}{\Tr{\rho(t_i\rightarrow-\infty)}}=\int \mathcal{D}(\phi^{cl},\phi^q) \prod_s\mathcal{D}(\theta_s^{cl},\theta_s^q)\ e^{iS[\phi,\theta_s]}=\int \mathcal{D}(\phi^{cl},\phi^q) \ e^{iS_\phi[\phi]}\int\prod_s\mathcal{D}(\theta_s^{cl},\theta_s^q)\ e^{i(S_\theta+S_{\phi\theta})},
    \end{align}
where the trace of the initial density matrix is taken into account in the normalization. Due to the action being quadratic it is possible to integrate out the $\theta_s$ fields and get an effective action for the $\phi$ photon field. In this case we must perform a Gaussian integral which results on the inverse of the matrix in Eq.~(\ref{freephotonprop}) and matrices coming from the coupling which has a $\tau_x$ Pauli Matrix. This means we can define:
\begin{align}
    \tilde{D}(\textbf{k},\omega)=-\tau_x \sum_s (t_s^{ph})^2 D_s(\textbf{k},\omega) \tau_x,\qquad
    D_s(\textbf{k},\omega) =  \begin{pmatrix}
    D_s^K(\textbf{k},\omega) && D_s^R(\textbf{k},\omega)\\
    D_s^A(\textbf{k},\omega) && 0
    \end{pmatrix},\qquad
    \tilde{D}(\textbf{k},\omega) =  \begin{pmatrix}
    0 && \tilde{D}^A(\textbf{k},\omega) \\ \tilde{D}^R(\textbf{k},\omega) &&
    \tilde{D}^K(\textbf{k},\omega) 
    \end{pmatrix}
\end{align}
where the previous matrix in the action is defined as the inverse of the propagator $D$ meaning that we have also the following relations:
\begin{align}
    D_s^{(R,A)}(\textbf{k},\omega)=\dfrac{1}{2 L}\dfrac{1}{(\omega\pm i0)^2-\nu_s^2(\textbf{k})}, \qquad D^{K}_s(\textbf{k},\omega)=\coth{\frac{\omega}{2T_{\text{cav}}}}(D^R_s(\textbf{k},\omega)-D^A_s(\textbf{k},\omega)),
\end{align}
where we have assumed that the bath is in thermal equilibrium so that the fluctuation-dissipation theorem relates the Keldysh Green's function to the retarder and advanced Green's functions. Focusing on the retarded and advanced Green's functions we obtain:
\begin{align}
    \tilde{D}^{(R,A)}(\textbf{k},\omega) = -\frac{1}{2 L}\sum_s \dfrac{(t_s^{ph})^2 L^2}{(\omega\pm i0)^2-\nu_s^2(\textbf{k})} = L \int \dfrac{d\tilde{\omega}}{2\pi}\dfrac{\tilde{\omega}J(\textbf{k},\tilde\omega)}{\tilde{\omega}^2-(\omega\pm i 0)^2}\qquad J(\textbf{k},\omega)= \pi \sum_s \dfrac{(t_s^{ph})^2}{\nu_s(\textbf{k})}\delta(\omega-\nu_s(\textbf{k})),
\end{align}
where we defined the bath spectral density as $J(\textbf{k},\omega)$ generically one can think of the spectral density of the bath as being some function of the frequency which is continuous in the real axis and decays faster than $1/\omega$ if thought as a complex function. In this case we can apply the  Sokhotski–Plemelj theorem which states for real functions that :
\begin{align}
    \lim _{\varepsilon \rightarrow 0^{+}} \int_a^b \frac{f(x)}{x \pm i \varepsilon} d x=\mp i \pi f(0)+\mathcal{P} \int_a^b \frac{f(x)}{x} d x.
\end{align}
In this case we can use it for $x=\tilde{\omega}^2-\omega^2$ and $\pm i \epsilon$ will be $\mp 2 i 0 \omega$ so that one obtains :
\begin{align}
    \tilde{D}^{(R,A)}(\textbf{k},\omega) = L \mathcal{P} \int \dfrac{dx}{2\pi}\dfrac{J(\textbf{k},x)}{x}\pm i 2 L \gamma_{ph}(\textbf{k},\omega)=cte.\pm i 2 L \gamma_{ph}(\textbf{k},\omega),
\end{align}
where we have defined $J$ so that one obtains also for the Keldysh Green's function the result:
\begin{align}
    \tilde{D}^{K}(\textbf{k},\omega)= 4 i\gamma_{ph}(\textbf{k},\omega) L B_0(\omega), \quad \gamma_{ph}(\textbf{k},\omega)=J(\textbf{k},\omega)/4.
\end{align}
In this way integrating out the bath variables leads to a contribution to the photon field action of the form:
\begin{align}
    \frac{1}{2}\sum_{\textbf{k}}\int_{-\infty}^{\infty} d\omega \Phi(\textbf{k},\omega)^\dagger \tilde{D}(\textbf{k},\omega)\Phi(\textbf{k},\omega).
\end{align}
The combined integration thus leads to an effective action which now has a finite Keldysh component as well as imaginary parts in the retarded and advanced Green's functions giving rise to the partition function of the form:
\begin{align}
    Z &=\int \mathcal{D}(\phi^{cl},\phi^q)\exp{i\sum_{\textbf{k}}\int_{-\infty}^{\infty} d\omega L \Phi^\dagger
    \begin{pmatrix}
    0 &&\omega^2-\nu(\textbf{k})^2- i  \gamma_{ph}(\textbf{k},\omega) \\
    \omega^2-\nu(\textbf{k})^2 + i  \gamma_{ph}(\textbf{k},\omega)&& 2 i\gamma_{ph}(\textbf{k},\omega)B_0(\omega)
    \end{pmatrix}
  \Phi},
\end{align}
where the constant change in the retarded and advanced functions was absorbed in the dispersion relation of the photon field.

\section{Electron field coupled to thermal bath}
  Analogously to the previous section we can consider the electrons coupled to a bath which physically could be the phonons of a lattice system. Since we consider only the generic change in behaviour to a thermal bath the may as well couple the electrons to another fermion system which will allow for analytically exact results in comparison to the phonon case but the overall generic behaviour. To test this we consider a fermion field with the corresponding action:
  
\begin{align}
    S_\psi=&\sum_{\textbf{k}}\oint_{\mathcal{C}}dt \ \bar{\psi}(\textbf{k},t)(i\partial_t-\epsilon(\textbf{k}))\psi(\textbf{k},t)=\sum_{\textbf{k}}\int_{-\infty}^{\infty}dt \  \bar{\psi}_+(\textbf{k},t)(i\partial_t-\epsilon(\textbf{k}))\psi_+(\textbf{k},t)- \bar{\psi}_-(\textbf{k},t)(i\partial_t-\epsilon(\textbf{k}))\psi_-(\textbf{k},t)\\
    &=\sum_{\textbf{k}}\int_{-\infty}^{\infty} d\omega \begin{pmatrix}
    \bar{\psi}^{cl}(\textbf{k},\omega) && \bar{\psi}^{q}(\textbf{k},\omega)
    \end{pmatrix}
    \begin{pmatrix}
    \omega-\epsilon(\textbf{k}) +i0 && 2 i 0 P(\omega) \\
    0 &&\omega-\epsilon(\textbf{k})-i0
    \end{pmatrix}
    \begin{pmatrix}
    \psi^{cl}(\textbf{k},\omega) \\ \psi^{q}(\textbf{k},\omega)
    \end{pmatrix},
\end{align}
 where the first line describes the field along the Keldysh contour being expressed in terms of the $+,-$ branches and the second line is Fourier transformed for each branch and a Keldysh rotation has been applied. For Fermions the Keldysh rotation is defined as $\psi^{(cl,q)}=(\psi_+\pm\psi_-)/\sqrt{2}$ and by convention $\bar{\psi}^{(cl,q)}=(\bar{\psi}_+\mp\bar{\psi}_-)/\sqrt{2}$. We can now think similarly to the previous case on a fermionic bath which is coupled to the previous electron field with the action:
\begin{align}
    S_\upsilon&=\sum_{\textbf{k},s}\int_{-\infty}^{\infty} d\omega \begin{pmatrix}
    \bar{\upsilon}_s^{cl}(\textbf{k},\omega) && \bar{\upsilon}_s^{q}(\textbf{k},\omega)
    \end{pmatrix}
    \begin{pmatrix}
    \omega-\epsilon_s(\textbf{k}) +i0 && 2 i 0 F_0(\omega) \\
    0 &&\omega-\epsilon_s(\textbf{k})-i0
    \end{pmatrix}
    \begin{pmatrix}
   \upsilon_s^{cl}(\textbf{k},\omega) \\ \upsilon_s^{q}(\textbf{k},\omega)
    \end{pmatrix},
\end{align}
 where $\upsilon_s^{q}(\textbf{k},\omega)$ are now the bath variables and $F_0(\omega)=\tanh((\omega-\mu_0)/2T_{\text{cry}})$ is the distribution function assuming the bath is in thermal equilibrium, which not necessarily implies the $\psi$ field is. To model the coupling between the bath and field we take the simplest interaction:
\begin{align}
    S_{\psi\upsilon}=\sum_{\textbf{k},s}t_s\oint_{\mathcal{C}}dt \  \bar{\psi}(\textbf{k},t) \upsilon_s(\textbf{k},t)+\bar{\upsilon}_s(\textbf{k},t) \psi(\textbf{k},t)=\sum_{\textbf{k},s}t_s\int_{-\infty}^{\infty} d\omega \ \bar{\Psi}(\textbf{k},\omega)^T\gamma^{cl}\Upsilon_s(\textbf{k},\omega)+ \bar{\Upsilon}_s(\textbf{k},\omega)^T\gamma^{cl}\Psi(\textbf{k},\omega),
\end{align}
 where the coupling changes for each field and the final vector fields are in Keldysh space meaning $\bar{\Psi}(\textbf{k},\omega)^T=\begin{pmatrix}
\bar{\psi}^{cl}(\textbf{k},\omega) && \bar{\psi}^{q}(\textbf{k},\omega)
\end{pmatrix}$ and $\Upsilon_s(\textbf{k},\omega)= \begin{pmatrix}
\eta_s^{cl}(\textbf{k},\omega) && \eta_s^{q}(\textbf{k},\omega)
\end{pmatrix}$. The matrix $\gamma^{cl}=\hat{1}$ is just the zeroth Pauli matrix in the Keldysh space. Now if one wants to obtain properties of the stationary state but arbitrary long time correlations then the Keldysh partition function is expressed as:

\begin{align}
    Z_\psi= \dfrac{\Tr{\rho(t_f\rightarrow\infty)}}{\Tr{\rho(t_i\rightarrow-\infty)}}=\int \mathcal{D}(\bar{\Psi},\Psi) \prod_s \mathcal{D}(\bar{\Upsilon}_s,\Upsilon_s)\ e^{iS[\Psi,\Upsilon_s]}=\int \mathcal{D}(\bar{\Psi},\Psi)\ e^{iS_\psi}\prod_s\mathcal{D}(\bar{\Upsilon}_s,\Upsilon_s)\ e^{i(S_\upsilon+S_{\psi\upsilon})},
\end{align}
 now since the action for the bath is again quadratic we can integrate them exactly to obtain an effective contribution to the $\psi$ field. In this case the integral of Grassman variables changes the form of the resulting integral so that it is useful to define:

\begin{align}
    \tilde{P    }(\textbf{k},\omega)=- \sum_s t_s^2 P_s(\textbf{k},\omega) ,\qquad
    P_s(\textbf{k},\omega) =  \begin{pmatrix}
     P_s^R(\textbf{k},\omega) && P_s^K(\textbf{k},\omega)\\
    0 && P_s^A(\textbf{k},\omega) 
    \end{pmatrix},
\end{align}
where the previous matrix in the action is defined as the inverse of the propagator $P$ meaning that we have also the following relations:
\begin{align}
    P_s^{(R,A)}(\textbf{k},\omega)=\dfrac{1}{\omega-\epsilon_s(\textbf{k})\pm i0}, \qquad P^{K}_s(\textbf{k},\omega)=\tanh(\dfrac{\omega-\mu_0}{2T_{\text{cry}}})(P^R_s(\textbf{k},\omega)-P^A_s(\textbf{k},\omega)),
\end{align}
where we have assumed that the bath is in thermal equilibrium so that the fluctuation-dissipation theorem applies. Focusing on the retarded and advanced Green's functions we obtain:
\begin{align}
    \tilde{P}^{(R,A)}(\textbf{k},\omega) = -\sum_s \dfrac{t_s^2}{\omega-\epsilon_s(\textbf{k})\pm i0} = \int \dfrac{d\tilde{\omega}}{2\pi}\dfrac{\rho(\textbf{k},\tilde\omega)}{\tilde{\omega}-\omega \mp i 0}\qquad \rho(\textbf{k},\omega)= 2\pi \sum_s t_s^2 \ \delta(\omega-\epsilon_s(\textbf{k})).
\end{align}
Applying again the Sokhotsky-Plemelj theorem with $x=\tilde{\omega}-\omega$ one obtains the result for the Greens function:

\begin{align}
    \tilde{P}^{(R,A)}(\textbf{k},\omega) = \mathcal{P} \int \dfrac{dx}{2\pi}\dfrac{\rho(\textbf{k},x)}{x}\pm i 2 \gamma(\textbf{k},\omega)=cte.\pm i 2 \gamma(\textbf{k},\omega),
\end{align}
where we have defined $\gamma$ so that one obtains also for the Keldysh Green's function the result:
\begin{align}
    \tilde{P}^{K}(\textbf{k},\omega)= 4 i\gamma(\textbf{k},\omega)F_0(\omega), \quad \gamma(\textbf{k},\omega)=\rho(\textbf{k},\omega)/4.
\end{align}
In this way integrating out the bath variables leads to a contribution to the electronic field action of the form:
\begin{align}
    \sum_{\textbf{k}}\int_{-\infty}^{\infty} d\omega \bar{\Psi}(\textbf{k},\omega)^T \tilde{P}(\textbf{k},\omega)\Psi(\textbf{k},\omega).
\end{align}
The combined integration thus leads to an effective action which now has a finite keldysh component as well as imaginary parts in the retarded and advanced Green's functions giving rise to the partition function of the form:
\begin{align}
    Z&=\int  \mathcal{D}(\bar{\Psi},\Psi)  \ e^{iS_{eff}[\psi]}
    &=\int  \mathcal{D}(\bar{\Psi},\Psi) \exp{i\sum_{\textbf{k}}\int_{-\infty}^{\infty} d\omega \bar{\Psi}^T
    \begin{pmatrix}
    \omega-\epsilon(\textbf{k})+ i  \gamma(\textbf{k},\omega) && 2 i\gamma(\textbf{k},\omega)F_0(\omega)\\
    0&&\omega-\epsilon(\textbf{k})- i  \gamma(\textbf{k},\omega) 
    \end{pmatrix}
  \Psi},
\end{align}
where the constant change in the retarded and advanced functions was absorbed in the dispersion relation of the photon field.

\section{Photon-Electron system and Dyson equation}

We are now interested in what happens to the two bath configuration in general when we add a coupling between the electrons and the photons, specifically specializing in the steady state resulting from such interactions. To address this we introduce a a Yukawa type interaction:
\begin{align}
    S_{int}&= g \oint_{\mathcal{C}} \dd t \int \dd^d\bm{x} \ \phi(\bm{x},t) \bar{\psi}(\bm{x},t)\psi(\bm{x},t) = g \int_{-\infty}^{\infty} \dd t \int \dd^d\bm{x} \ \phi_+(\bm{x},t) \bar{\psi}_+(\bm{x},t)\psi_+(\bm{x},t)-\phi_-(\bm{x},t) \bar{\psi}_-(\bm{x},t)\psi_-(\bm{x},t)\\
    &=g \int_{-\infty}^{\infty} \dd t \int \dd^d\bm{x} \ \ \Phi_\alpha(\bm{x},t) \bar{\Psi}_a(\bm{x},t) \gamma^{\alpha}_{ab}\Psi_b(\bm{x},t), \qquad \gamma^{cl}=\begin{pmatrix}
    1 && 0 \\
    0 && 1
    \end{pmatrix},\quad \gamma^{q}=\begin{pmatrix}
    0 && 1 \\
    1 && 0
    \end{pmatrix}.
\end{align}
A good check is to see what the units of the coupling are by using the fact that previously all fields in time had dimensionless units which means for the interacting part of the action to be dimensionless we must require $g$ to have units of energy. In cavities the coupling has indeed units of energy and is typically of the order $0.1t$ for a hopping parameter $t$ of the electrons. If one is interested in the form of the Green's function for long times meaning analyzing properties of the steady state one will generally accumulate errors in time if one uses a perturbative approach. It is in this spirit that non-perturbative approaches such as the self-consistent formulation in terms of the effective action lead to information not being lost specifically conserved quantities. It is in this spirit that one defines a quantity  $\Gamma$ which is gauge invariant and respects the microscopic symmetries from which one can obtain the self-energy of the system. To derive the corresponding equations is useful to define the following Feynman diagrams:

\begin{align}
    &\begin{fmffile}{1}
i G_{ab}(x,x')
    =\hspace{1cm}\begin{gathered}
 \begin{fmfgraph*}(60,40)
 \fmfstraight
   \fmfleft{i1}
   \fmfright{o1}
   \fmf{dbl_plain_arrow}{i1,o1}
   \fmflabel{$b,x'$}{i1}
   \fmflabel{$a,x$}{o1}
 \end{fmfgraph*}
\end{gathered}
\end{fmffile}\hspace{1cm}
=  \int  \mathcal{D}(\bar{\Psi},\Psi) \int \mathcal{D}\Phi \  e^{iS[\Psi,\Phi]} \Psi_a(x)\bar{\Psi}_b(x') =\expval{\Psi_a(x)\bar{\Psi}_b(x') } \\
&\begin{fmffile}{2}
i G^0_{ab}(x,x')
    =\hspace{1cm}\begin{gathered}
 \begin{fmfgraph*}(60,40)
 \fmfstraight
   \fmfleft{i1}
   \fmfright{o1}
   \fmf{fermion}{i1,o1}
   \fmflabel{$b,x'$}{i1}
   \fmflabel{$a,x$}{o1}
 \end{fmfgraph*}
\end{gathered}
\end{fmffile}\hspace{1cm}
=  \expval{\Psi_a(x)\bar{\Psi}_b(x')}_0\\
& \begin{fmffile}{3}
 iU_{\alpha\beta}(x,x')
    =\hspace{1cm}\begin{gathered}
 \begin{fmfgraph*}(60,40)
 \fmfstraight
   \fmfleft{i1}
   \fmfright{o1}
   \fmf{dbl_wiggly}{i1,o1}
   \fmflabel{$\beta,x'$}{i1}
   \fmflabel{$\alpha,x$}{o1}
 \end{fmfgraph*}
\end{gathered}
\end{fmffile}\hspace{1cm}= \int  \mathcal{D}(\bar{\Psi},\Psi) \int \mathcal{D}\Phi \  e^{iS[\Psi,\Phi]} \Phi_\alpha(x)\Phi_\beta(x') =\expval{ \Phi_\alpha(x)\Phi_\beta(x')} \\
&\begin{fmffile}{4}
 iU^0_{\alpha\beta}(x,x')
    =\hspace{1cm}\begin{gathered}
 \begin{fmfgraph*}(60,40)
 \fmfstraight
   \fmfleft{i1}
   \fmfright{o1}
   \fmf{boson}{i1,o1}
   \fmflabel{$\beta,x'$}{i1}
   \fmflabel{$\alpha,x$}{o1}
 \end{fmfgraph*}
\end{gathered}
\end{fmffile}\hspace{1cm}=\expval{\Phi_\alpha(x)\Phi_\beta(x')}_0\\
&\hspace{3cm}\begin{fmffile}{5}\begin{gathered}
 \begin{fmfgraph*}(60,40)
   \fmfleft{i1,o1}
   \fmfright{o2}
   \fmf{plain}{i1,v1,o1}
   \fmf{boson}{v1,o2}
   \fmflabel{$b$}{i1}
   \fmflabel{$a$}{o1}
   \fmflabel{$\alpha$}{o2}
   \fmflabel{$x$}{v1}
 \end{fmfgraph*}
\end{gathered}
\end{fmffile}\hspace{1cm}=i g \gamma^{\alpha}_{ab},
\end{align}
 where the expectation values in the path integral language are already path ordered in the operator language over the Keldysh contour and the last diagram represents the interaction vertex where the plain straight lines are the end and beginning of the propagators only. In the conserving approximation one defines an effective action in terms of bubble diagrams which have no external legs and defines the self-energy to be the functional derivative with respect to the propagators:
\begin{align}
      \Sigma^\psi_{ab}(x,x')=\dfrac{\delta\Gamma}{\delta G_{ba}(x',x)} \qquad
      \Sigma^\phi_{\alpha\beta}(x,x')=\dfrac{\delta\Gamma}{\delta U_{\beta\alpha}(x',x)}.
\end{align}
The previous relation holds up to a combinatorial factor that counts the number of equivalent diagrams(the factor is $1/n$ for $n$ vertices) and a $\pm$ sign for photons or electrons.We can now define the effective action in terms of the fully dressed propagators to first order in the photon propagator:\\

\begin{align}
\begin{fmffile}{a1}
    \Gamma^{HF}=\hspace{1cm}\begin{gathered}
 \begin{fmfgraph}(120,80)
   \fmfleft{v1}
   \fmfright{v4}
   \fmf{dbl_plain_arrow,left}{v1,v2}
   \fmf{dbl_plain_arrow,left}{v2,v1}
   \fmf{boson}{v2,v3}
   \fmf{dbl_plain_arrow,left}{v3,v4}
   \fmf{dbl_plain_arrow,left}{v4,v3}
 \end{fmfgraph}
\end{gathered}
\end{fmffile}\hspace{0.5cm}
+\hspace{0.5cm}
\begin{fmffile}{a2}\begin{gathered}
 \begin{fmfgraph}(50,30)
   \fmfleft{v1}
   \fmfright{v2}
   \fmf{dbl_wiggly}{v1,v2}
   \fmffreeze
   \fmf{dbl_plain_arrow,left}{v1,v2}
   \fmf{dbl_plain_arrow,left}{v2,v1}
 \end{fmfgraph}
\end{gathered}
\end{fmffile}\hspace{0.5cm} +\hspace{0.5cm} \dots
\end{align}
In this approximation we obtain that the self-energy is given by:
\begin{align}
      \Sigma^\psi_{ab}(x,x')=-ig^2\delta(x-x')\int \dd y \ \gamma^\alpha_{a b} \gamma^{\alpha'}_{a'b'}U_{\alpha\alpha'}^0(x,y)G_{b'a'}(y,y)+ig^2\gamma^\alpha_{a b'} \gamma^{\alpha'}_{a'b}U_{\alpha\alpha'}(x,x')G_{b'a'}(x,x'),
\end{align}
while for the photon to lowest order the first term (Hartree) doesn't contribute since it has no dressed propagator and the second one gives the bubble diagram self energy:
\begin{align}
      \Sigma^\phi_{ab}(x,x')= -ig^2\gamma^\alpha_{a b} \gamma^{\beta}_{b'a'}G_{bb'}(x,x')G_{a'a}(x',x).
\end{align}

We can make use of the self-energy to obtain steady-state behaviour by substituting and solving self-consistently in the Dyson equation:

\begin{align}
\begin{fmffile}{Dyson1}\begin{gathered}
 \begin{fmfgraph*}(60,40)
 \fmfstraight
   \fmfleft{i1}
   \fmfright{o1}
   \fmf{dbl_plain_arrow}{i1,o1}
 \end{fmfgraph*}
\end{gathered}
\end{fmffile}\hspace{0.5cm}
=\hspace{0.5cm}
\begin{fmffile}{Dyson2}\begin{gathered}
 \begin{fmfgraph*}(60,40)
 \fmfstraight
   \fmfleft{i1}
   \fmfright{o1}
   \fmf{fermion}{i1,o1}
 \end{fmfgraph*}
\end{gathered}
\end{fmffile}\hspace{0.5cm}
+\hspace{0.5cm}
\begin{fmffile}{Dyson3}\begin{gathered}
\Large \color{white}
 \begin{fmfgraph*}(100,40)
 \fmfstraight
   \fmfleft{i1}\fmfright{o1}
   \fmfblob{1cm}{v1}
   \fmf{fermion}{i1,v1}
   \fmf{dbl_plain_arrow}{v1,o1}
 \end{fmfgraph*}
\end{gathered}
\end{fmffile},
\end{align}
where the shaded circle represents the self-energy $\Sigma$. When written in terms of the Keldysh components it becomes:
\begin{align}
    &((G_0^{R/A})^{-1}-\Sigma^{R/A}_\psi)\circ G^{R/A}=\delta(x-x'), \qquad G^K=G^R\circ(-(G_0^{K})^{-1}+\Sigma^K_\psi)\circ G^A\\
    &((U_0^{R/A})^{-1}-\Sigma^{R/A}_\phi)\circ U^{R/A}=\delta(x-x'), \qquad U^K=G^R\circ(-(U_0^{K})^{-1}+\Sigma^K_\phi)\circ U^A.
\end{align}
Now it is common to reparameterize the anti-hermitian Keldysh Green's function in terms of a hermitian function $F$ defined by $G^K=G^R\circ F -F\circ G^A$. Using the previous Dyson equations one obtains:
\begin{align}
    (G_0^K)^{-1}&=G_R^{-1}\circ F-F\circ G_A^{-1}+\Sigma^K\\
    &=((G_0^R)^{-1}-\Sigma^R)\circ F-F\circ ((G_0^A)^{-1}-\Sigma^A))+\Sigma^K\\
    &=\Sigma^K-(\Sigma^R\circ F-F\circ \Sigma^A)+((G_0^R)^{-1}\circ F-F\circ (G_0^A)^{-1}).
\end{align}

Assuming stationarity or time translation invariance and space translation invariance while using the identities $(G^R)^\dagger=G^A$, $(G^K)^\dagger=-G^K$ we can obtain the quantum kinetic equations (QKE):
\begin{align}
    & 2i\gamma(\textbf{k},\omega)(F_0(\omega)-F(\textbf{k},\omega))=\Sigma^K_\psi(\textbf{k},\omega)-2i F(\textbf{k},\omega)\Im \Sigma^R_\psi(\textbf{k},\omega)\\
    &2i\gamma_{ph}(\textbf{k},\omega)(B_0(\omega)-D(\textbf{k},\omega))=\Sigma^K_\phi(\textbf{k},\omega)-2i D(\textbf{k},\omega)\Im \Sigma_\phi^R(\textbf{k},\omega),
\end{align}
where we used the previously derived non-interacting Green's function obtained from integrating out the bath degrees of freedom. Since the bath gave a generic tunable function $F_0$ this will not generically equal the thermal $F_{th}$ which means the right side of the QKE is not zero meaning the steady state can be nonthermal. Written explicitly the Keldysh and retarded self-energies of the electrons one obtains:
\begin{align}
    \Sigma^K_\psi(x,x')=ig^2(G^K(x,x') U^K(x',x)-(G^R(x,x')-G^A(x,x'))(U^R(x',x)-U^A(x',x)))\\
    \Sigma^R_\psi(x,x')=-ig^2\delta(x-x')\int \dd y U_0^R(x,y)G^K(y,y)+ig^2(G^R(x,x')U^K(x',x)+G^K(x,x')U^A(x',x)),
\end{align}
where we used the causality identities $G^R(x,y)U^R(y,x)=0,\ G^R(x,x)+G^A(x,x)=0$ and others similar. 
We now take as a first approximation that the coupling $g$ between photons and electrons is very small compared to the dissipative terms of the baths $\gamma,\gamma_{ph}$ and the photon cavity frequency $\nu_0$ so that we can use the retarded and advanced functions of the non-interacting limit. Let us define the momentum Fourier transform of the fields as in Eq.~(5.8) of the Kamenev book \cite{kamenev2023field}, $\phi(\textbf{x})=\sum_{\textbf{k}}\phi(\textbf{k})e^{i\textbf{k}\cdot \textbf{x}}$. So that $U^K_0(\textbf{k},\omega)=B_0(\omega)(U^R_0(\textbf{k},\omega)-U^A_0(\textbf{k},\omega))$ is satisfied and we can obtain:
\begin{align}
    &\Sigma^K_\psi(\textbf{k},\omega)=i g^2 \sum_{\textbf{k}'}\int \dfrac{\dd \tilde{\omega} }{2\pi}(G^R(\textbf{k}',\tilde{\omega})-G^A(\textbf{k}',\tilde{\omega}))(U^R(\textbf{k}'-\textbf{k},\tilde{\omega}-\omega)-U^A(\textbf{k}'-\textbf{k},\tilde{\omega}-\omega))(F(\textbf{k}',\tilde{\omega})B_0(\tilde{\omega}-\omega)-1)\\
    &\Sigma^R_\psi(\textbf{k},\omega)-\Sigma^A_\psi(\textbf{k},\omega)=i g^2\sum_{\textbf{k}'}\int   \dfrac{\dd \tilde{\omega} }{2\pi}(G^R(\textbf{k}',\tilde{\omega})-G^A(\textbf{k}',\tilde{\omega}))(U^R(\textbf{k}'-\textbf{k},\tilde{\omega}-\omega)-U^A(\textbf{k}'-\textbf{k},\tilde{\omega}-\omega))\notag\\
   & \hspace{11cm} \times (B_0(\tilde{\omega}-\omega)-F(\textbf{k}',\tilde{\omega})),
\end{align}
the corresponding QKE for the electrons becomes:
\begin{align}
    2i\gamma(\textbf{k},\omega)(F_0(\omega)-F(\textbf{k},\omega))=ig^2 \sum_{\textbf{k}'}\int  \dfrac{\dd \tilde{\omega} }{2\pi}(G^R_0(\textbf{k}',\tilde{\omega})-G^A_0(\textbf{k}',\tilde{\omega}))(U^R_0(\textbf{k}'-\textbf{k},\tilde{\omega}-\omega)-U^A_0(\textbf{k}'-\textbf{k},\tilde{\omega}-\omega))&\notag\\ 
    (F(\textbf{k}',\tilde{\omega})B_0(\tilde{\omega}-\omega)-1-F(\textbf{k},\omega)(B_0(\tilde{\omega}-\omega)-F(\textbf{k}',\tilde{\omega})).&
\end{align}

\section{Electron-bath coupling bigger than photon-bath: \texorpdfstring{$\gamma \gg \gamma_{ph}$}{gamma}}
In this regime we have that $\gamma \gg \gamma_{ph}$ and both satisfy 
,$\gamma_{ph},\gamma \gg g^2$ in order to substitute the original Green's function for the photons and electrons. Let's now define the function 
\begin{align}
    H(\tilde{\omega})=\tilde{H}(\omega,\tilde{\omega};\textbf{k},\textbf{k}')=     F(\textbf{k}',\tilde{\omega})B_0(\tilde{\omega}-\omega)-F(\textbf{k},\omega)(B_0(\tilde{\omega}-\omega)-F(\textbf{k}',\tilde{\omega}))-1,
\end{align}
taking the photons to describe cavity photons we have then a cavity frequency $\nu_0$ and the speed of light coming inside the dispersion:
\begin{align}
    \nu(\textbf{k})^2=\nu_0^2+c^2\abs{\textbf{k}}^2,
\end{align}
 under the assumption that $c$ is the biggest scale and we are interested on long time physics we see that the relevant length scale of variation of the photons going like $ct$ is much larger than any material length scale and as such we can take the propagator to be constant on those length scales which in momentum space implies :
\begin{align}
    &(U^R_0(\textbf{k}'-\textbf{k},\tilde{\omega}-\omega)-U^A_0(\textbf{k}'-\textbf{k},\tilde{\omega}-\omega))= i  \delta_{\textbf{k},\textbf{k}'}\dfrac{2}{2L}\Im \dfrac{1}{(\tilde{\omega}-\omega)^2-\nu_0^2+ i  \gamma_{ph}(\tilde{\omega}-\omega)}\label{distfuncU}\\
    &=i \dfrac{1}{L} \delta_{\textbf{k},\textbf{k}'} \dfrac{-\gamma_{ph}(\tilde{\omega}-\omega)}{((\tilde{\omega}-\omega)^2-\nu_0^2)^2+  \gamma_{ph}^2(\tilde{\omega}-\omega)}\approx -i\dfrac{1}{L}  \pi\   \delta_{\textbf{k},\textbf{k}'}\delta((\tilde{\omega}-\omega)^2-\nu_0^2) \text{sign}(\tilde{\omega}-\omega),
\end{align}
where we assumed $\gamma_{ph}(\omega)=\gamma_{ph} \omega$. Let us use units of $c=1$ as mentioned in the main text. The spectral function of the electrons would be a lorentzian:
\begin{align}
    &(G^R_0(\textbf{k}',\tilde{\omega})-G^A_0(\textbf{k}',\tilde{\omega}))=2i \dfrac{-\gamma(\textbf{k}',\tilde{\omega})}{(\tilde{\omega}-\epsilon_{\textbf{k}'})^2+  \gamma^2(\textbf{k}',\tilde{\omega})}.
\end{align}
We can now group everything inside the QKE to find:
\begin{align}
    &2i\gamma(\textbf{k},\omega)(F_0(\omega)-F(\textbf{k},\omega))=i\dfrac{g^2}{L} \sum_{\textbf{k}'} \int  \dfrac{\dd \tilde{\omega} }{2\pi} 2i \dfrac{\gamma(\textbf{k}',\tilde{\omega})\text{sign}(\tilde{\omega}-\omega)}{(\tilde{\omega}-\epsilon_{\textbf{k}'})^2+  \gamma^2(\textbf{k}',\tilde{\omega})} i \pi \delta_{\textbf{k},\textbf{k}'}\delta((\tilde{\omega}-\omega)^2-\nu_0^2) 
    H(\tilde{\omega})\\
    &\gamma(\textbf{k},\omega)(F_0(\omega)-F(\textbf{k},\omega))=-\pi\dfrac{g^2}{L} \sum_{\textbf{k}'}\dfrac{\dd \tilde{\omega} }{2\pi} \dfrac{\gamma(\textbf{k}',\tilde{\omega})\text{sign}(\tilde{\omega}-\omega)}{(\tilde{\omega}-\epsilon_{\textbf{k}'})^2+  \gamma^2(\textbf{k}',\tilde{\omega})} \delta_{\textbf{k},\textbf{k}'}\delta((\tilde{\omega}-\omega)^2-\nu_0^2)H(\tilde{\omega}),
\end{align}
the delta function in frequency can equivalently be written as:
\begin{align}
    \text{sign}(\tilde{\omega}-\omega)\delta((\tilde{\omega}-\omega)^2-\nu_0^2)=\dfrac{1}{2\abs{\nu_0}}(\delta(\tilde{\omega}-(\omega+\nu_0))-\delta(\tilde{\omega}-(\omega-\nu_0))),
\end{align}
meanwhile the sum in momentum together with the delta yields one. We then obtain:
\begin{align}
    (F_0(\omega)-F(\textbf{k},\omega))=&-\dfrac{ g^2 }{4L\abs{\nu_0}\gamma(\textbf{k},\omega)}\left(\dfrac{\gamma(\textbf{k},\omega+\nu_0)H(\omega+\nu_0)}{(\omega+\nu_0-\epsilon_{\textbf{k}})^2+  \gamma^2(\textbf{k},\omega+\nu_0)}-\dfrac{\gamma(\textbf{k},\omega-\nu_0)H(\omega-\nu_0)}{(\omega-\nu_0-\epsilon_{\textbf{k}})^2+  \gamma^2(\textbf{k},\omega-\nu_0)}\right)\\
    &=A_{\nu_0}(\textbf{k},\omega)H(\omega+\nu_0)-A_{-\nu_0}(\textbf{k},\omega)H(\omega-\nu_0),
\end{align}
where we have defined the function:
\begin{align}
 A_{\nu}(\textbf{k},\omega)=-\dfrac{ g^2 }{4\pi \gamma(\textbf{k},\omega)}\dfrac{\gamma(\textbf{k},\omega+\nu)}{(\omega+\nu-\epsilon_{\textbf{k}})^2+  \gamma^2(\textbf{k},\omega+\nu)}.
\end{align}

We can now write explicitly the function $H(\omega)$ and collect terms which depend on the distribution function $F(\textbf{k},\omega)$ so that we have:
\begin{align}
 F_0(\omega)-B_0 A_{\nu_0}(\textbf{k},\omega)F(\textbf{k},\omega+\nu_0)&-B_0 A_{-\nu_0}(\textbf{k},\omega)F(\textbf{k},\omega-\nu_0)-A_{-\nu_0}(\textbf{k},\omega)+A_{\nu_0}(\textbf{k},\omega)=F(\textbf{k},\omega)\left(1-B_0 A_{\nu_0}(\textbf{k},\omega)\right. \notag \\
&\left.-B_0 A_{-\nu_0}(\textbf{k},\omega)+ A_{\nu_0}(\textbf{k},\omega)F(\textbf{k},\omega+\nu_0)-A_{-\nu_0}(\textbf{k},\omega)F(\textbf{k},\omega-\nu_0)\right),
\end{align}
where we defined $B_0=B_0(\nu_0)$ and used that $B_0(-\nu_0)=-B_0(\nu_0)$. We can now rewrite the distribution function as a function of the other parameters and itself at another frequency meaning:
\begin{align}
  F(\textbf{k},\omega)= \dfrac{F_0(\omega)+A_{\nu_0}(\textbf{k},\omega)(1-B_0F(\textbf{k},\omega+\nu_0))-A_{-\nu_0}(\textbf{k},\omega)(1+B_0F(\textbf{k},\omega-\nu_0))}{1+A_{\nu_0}(\textbf{k},\omega)(F(\textbf{k},\omega+\nu_0)-B_0)-A_{-\nu_0}(\textbf{k},\omega)(F(\textbf{k},\omega-\nu_0)+B_0)}.
\end{align}
Let us first treat the case of $\nu_0\gg 0$ in this limit we obtain essentially the equilibrium function since $A_{\nu_0}\propto 1/\nu_0^2$ so $F(\textbf{k},\omega)\approx F_0(\omega)$ to lowest order in the frequency. The case of $\nu_0 \approx 0$ is more interesting since in this case we can expand around zero to obtain, 
 expanding in Mathematica to first order:

\begin{align}
    \lim_{\nu_0\rightarrow 0}F(\textbf{k},\omega)&=\dfrac{F_0(\omega) -\frac{4T_{\text{cav}}}{\nu_0}A_0F+2\nu_0\left(-\frac{1}{6T_{\text{cav}}}A_0F-T_{\text{cav}} A_0\dfrac{\partial^2 F(\textbf{k},\omega)}{\partial \omega^2 }+A'_0-2T_{\text{cav}} A'_0\dfrac{\partial F(\textbf{k},\omega)}{\partial \omega }-T_{\text{cav}} A''_0F\right)}{1-\frac{4T_{\text{cav}}}{\nu_0}A_0-2\nu_0\left(\frac{1}{6T_{\text{cav}}}A_0-A_0\dfrac{\partial F(\textbf{k},\omega)}{\partial \omega }-A'_0F+T_{\text{cav}} A''_0\right)},\\
    F(\textbf{k},\omega)-F_0(\omega)&=2\nu_0\left[A'_0\left(1-F(\textbf{k},\omega)^2\right)-\left(A_0F(\textbf{k},\omega)+2T_{\text{cav}}A'_0\right)\dfrac{\partial F(\textbf{k},\omega)}{\partial \omega }-T_{\text{cav}} A_0\dfrac{\partial^2 F(\textbf{k},\omega)}{\partial \omega^2 }\right]~~\text{where},\\  A_0&=-\dfrac{g^2 }{4\pi}\dfrac{1}{(\omega-\epsilon_{\textbf{k}})^2+  \gamma^2(\textbf{k},\omega)}~~\text{and}~~A'_0=\dfrac{ g^2 }{4\pi}\dfrac{2(\omega-\epsilon_{\textbf{k}})+2\gamma(\textbf{k},\omega)\frac{\partial \gamma}{\partial \omega}}{[(\omega-\epsilon_{\textbf{k}})^2+  \gamma^2(\textbf{k},\omega)]^2}.
\end{align}
in a simplified form we can write this as:
\begin{align}
    \Delta F=-\dfrac{g^2\nu_0 \gamma(\textbf{k},\omega)}{4\pi A(\textbf{k},\omega)}\dfrac{\partial}{\partial \omega}\left(\dfrac{A(\textbf{k},\omega)^2}{\gamma(\textbf{k},\omega)^2}\left(1-F(\textbf{k},\omega)^2-2T_{\text{cav}}\dfrac{\partial}{\partial \omega}F(\textbf{k},\omega)\right)\right), \label{Fdiff_QKE}
\end{align}
let's examine the limit $g\rightarrow\infty$. In that case we will get the following equation for $F(\bk,\om)$
\beq\label{eqFlargeg}
\tilde{A}'_0\left(1-F^2\right)-\left(F+2\tm \tilde{A}'_0\right)\dF-\tm \ddF=0, \quad \tilde{A}'_0=\dfrac{2(\omega-\epsilon_{\textbf{k}})+2\gamma(\textbf{k},\omega)\frac{\partial \gamma}{\partial \omega}}{(\omega-\epsilon_{\textbf{k}})^2+  \gamma^2(\textbf{k},\omega)}
\eeq
In this limit we would expect that the fermions would thermalize to the photon bath with the temperature $\tm$. To check that let's assume that
\beq
F(\bk,\om)=\tanh\left(\frac{\omega-\tilde{\mu}}{2\tm}\right).
\eeq
So, we have 
\beq
\dF=\frac{1}{2\tm}\left[\frac{1}{\cosh\left(\frac{\omega-\tilde{\mu}}{2\tm}\right)}\right]^2
\eeq
and 
\beq
\begin{split}
\ddF&=\frac{1}{2\tm}(-2)\left[\frac{1}{\cosh\left(\frac{\omega-\tilde{\mu}}{2\tm}\right)}\right]^2\tanh\left(\frac{\omega-\tilde{\mu}}{2\tm}\right)\frac{1}{2\tm}\\
&=-\frac{1}{2\tm^2}\left[\frac{1}{\cosh\left(\frac{\omega-\tilde{\mu}}{2\tm}\right)}\right]^2\tanh\left(\frac{\omega-\tilde{\mu}}{2\tm}\right)
\end{split}
\eeq
Using this on the left hand side of Eq.~(\ref{eqFlargeg}), we get 
\beq
\begin{split}
&\tilde{A}'_0\left(1-F^2\right)-\left(F+2\tm \tilde{A}'_0\right)\dF-\tm \ddF\\
&=\tilde{A}'_0\left[1-\tanh^2\left(\frac{\omega-\tilde{\mu}}{2\tm}\right)\right]-\left(F+2\tm \tilde{A}'_0\right)\frac{1}{2\tm}\left[\frac{1}{\cosh\left(\frac{\omega-\tilde{\mu}}{2\tm}\right)}\right]^2\\
&-\tm\left(-\frac{1}{2\tm^2}\right)\left[\frac{1}{\cosh\left(\frac{\omega-\tilde{\mu}}{2\tm}\right)}\right]^2\tanh\left(\frac{\omega-\tilde{\mu}}{2\tm}\right)\\
&=\tilde{A}'_0\left[\frac{1}{\cosh\left(\frac{\omega-\tilde{\mu}}{2\tm}\right)}\right]^2-\frac{1}{2\tm}\left[\frac{1}{\cosh\left(\frac{\omega-\tilde{\mu}}{2\tm}\right)}\right]^2\tanh\left(\frac{\omega-\tilde{\mu}}{2\tm}\right)-\tilde{A}'_0\left[\frac{1}{\cosh\left(\frac{\omega-\tilde{\mu}}{2\tm}\right)}\right]^2\\
&+\frac{1}{2\tm}\left[\frac{1}{\cosh\left(\frac{\omega-\tilde{\mu}}{2\tm}\right)}\right]^2\tanh\left(\frac{\omega-\tilde{\mu}}{2\tm}\right)\\
&=0
\end{split}
\eeq
So, $F(\bk,\om)=\tanh\left(\frac{\omega-\tilde{\mu}}{2\tm}\right)$ solves Eq.~(\ref{eqFlargeg}). Indeed in the $g\rightarrow\infty$ limit the fermions thermalize at the photon temperature as expected. 
\section{Approximate solution for \texorpdfstring{$g^2\nu_0/\gamma^3\rightarrow 0$}{gnu}}

Let us solve this equation for small values of $\lambda=\dfrac{g^2\nu_0 }{2\pi\gamma_0^3}$ which means we expand the solution of the differential equation as:
\begin{align}
    F(\textbf{k},\omega)=f_0(\textbf{k},\omega)+\lambda f_1(\textbf{k},\omega)+\lambda^2 f_2(\textbf{k},\omega)+\cdots
\end{align}
We specialize here to the simplified differential equation with $\gamma(\textbf{k},\omega)=\gamma_0$, let us measure all energies, temperatures and frequencies in terms of $\gamma_0$ meaning $\omega\rightarrow \gamma_0 \omega, \epsilon_\textbf{k}\rightarrow \gamma_0 \epsilon_\textbf{k},\dots $ so that we have the equation:
\begin{align}
    F(\textbf{k},\omega)-F_0(\omega)&=\lambda B(\textbf{k},\omega)\left[\tilde{A}(\textbf{k},\omega)\left(1-F(\textbf{k},\omega)^2\right)+\left(F(\textbf{k},\omega)-2T_{\text{cav}}\tilde{A}(\textbf{k},\omega)\right)\dfrac{\partial F(\textbf{k},\omega)}{\partial \omega }+T_{\text{cav}} \dfrac{\partial^2 F(\textbf{k},\omega)}{\partial \omega^2 }\right],
\end{align}
where
\begin{align}
    B(\textbf{k},\omega)&=\dfrac{1}{(\omega-\epsilon_{\textbf{k}})^2+ 1},\quad \tilde{A}(\textbf{k},\omega)=2\dfrac{\omega-\epsilon_{\textbf{k}}}{(\omega-\epsilon_{\textbf{k}})^2+1}=2(\omega-\epsilon_{\textbf{k}}) B(\textbf{k},\omega)
\end{align}
We see that the right hand side is already linear in $\lambda$ even to the zeroth order of the distribution function which means that:
\begin{align}
    f_0(\textbf{k},\omega)=F_0(\omega)
\end{align}
Let us simplify further by changing variables to $x=\omega-\epsilon_{\textbf{k}}$ and suppress the momentum dependence so the equation simplifies to:
\begin{align}
    F(x)-f_0(x)= \lambda \dfrac{1}{x^2+1}\left(\dfrac{2x}{x^2+1}(1-F(x)^2)+\left(F(x)-2T_{\text{cav}}\dfrac{2x}{x^2+1}\right)\dfrac{\dd F}{\dd x}+T_{\text{cav}} \dfrac{\dd^2 F}{\dd x^2}\right)
\end{align} 
to linear order we now have:
\begin{align}
    \lambda f_1(x)=\lambda \dfrac{1}{x^2+1}\left(\dfrac{2x}{x^2+1}(1-f_0(x)^2)+f_0(x)f_0'(x)-2T_{\text{cav}}\dfrac{2x}{x^2+1}f_0'(x)+T_{\text{cav}} \dfrac{\dd^2 f_0}{\dd x^2}\right),
\end{align}
where we have defined $f_0(x)=\tanh(\dfrac{x+\epsilon_{\textbf{k}}-\mu_0}{2T_{\text{cry}}})$, substituting and simplifying we obtain to linear order in $\lambda$:
\begin{align}
    F(\textbf{k},\omega)=F_0(\omega)+\tilde{g}\frac{\left(T_{\text{cry}}-T_{\text{cav}}\right) \left(\left((\omega-\epsilon_{\textbf{k}})^2+1\right) \tanh
   \left(\frac{\omega-\mu _0}{2 T_{\text{cry}}}\right)+4 T_{\text{cry}} (\omega-\epsilon_{\textbf{k}}) \right)
   \text{sech}^2\left(\frac{\omega-\mu _0}{2 T_{\text{cry}}}\right)}{2
   \left((\omega-\epsilon_{\textbf{k}})^2+1\right)^2 T_{\text{cry}}^2}
\end{align}
, where we defined $\tilde{g}=\dfrac{ g^2\nu_0 }{2\pi\gamma_0^3}$. We can now evaluate this for $\omega=\epsilon_{\textbf{k}}$ to obtain:
\begin{align}
    F(\textbf{k},\epsilon_{\textbf{k}})=\tanh \left(\frac{\mu_0-\epsilon_{\textbf{k}}}{2
T_{\text{cry}}}\right)+\dfrac{(T_{\text{cry}}-T_{\text{cav}})\tilde{g}}{2 T_{\text{cry}}^2}\text{sech}^2\left(\frac{\mu_ 0-\epsilon_{\textbf{k}}}{2 T_{\text{cry}}}\right)
 \tanh \left(\frac{\mu_0-\epsilon_{\textbf{k}}}{2
T_{\text{cry}}}\right)
\end{align}
Putting back the $\gamma$ dependence by reversing the previous transformation leads to:
\begin{align}
    \Delta F = \dfrac{g^2\nu_0 }{2\pi}\frac{\left(T_{\text{cry}}-T_{\text{cav}}\right) \text{sech}^2\left(\frac{\omega -\mu _0}{2
   T_{\text{cry}}}\right)\left(4T_{\text{cry}} (\omega -\epsilon_{\textbf{k}}) +\left(\gamma_0 ^2+(\omega -\epsilon_{\textbf{k}})^2\right) \text{tanh}\left(\frac{\omega -\mu _0}{2T_{\text{cry}}}\right)\right)}{2
   \left((\omega-\epsilon_{\textbf{k}})^2+\gamma_0^2\right)^2 T_{\text{cry}}^2}
\end{align}
We analyze the next higher order correction in $\nu_0$ to the distribution function which gives a contribution to $f_1$ of the form, for $T_{\text{cav}}\rightarrow 0$ to lowest order:
\begin{align}
    \Delta F^{(3)}=-\dfrac{ g^2\nu_0^3 }{2\pi}\frac{\text{sech}^2\left(\frac{\omega -\mu _0}{2
   T_{\text{cry}}}\right)\left(4T_{\text{cry}} (\omega -\epsilon_{\textbf{k}}) +\left(\gamma_0 ^2+(\omega -\epsilon_{\textbf{k}})^2\right) \text{tanh}\left(\frac{\omega -\mu _0}{2T_{\text{cry}}}\right)\right)}{12 T_{\text{cry}}^2
   T_{\text{cav}} \left(\gamma_0 ^2+(\omega -\epsilon_{\textbf{k}})^2\right)^2}
\end{align}
The complete solution correct to second order in $\tilde{g}$ and sixth order in frequency $\nu_0$ is,where $\Delta T = T_{\text{cav}}-T_{\text{cry}}$: 
\begin{align}
    \Delta F(\textbf{k},\omega) = -\dfrac{ g^2\nu_0(\Delta T+\nu_0^2/(6T_{\text{cav}})) }{4\pi}\frac{\text{sech}^2\left(\frac{\omega -\mu _0}{2
   T_{\text{cry}}}\right)\left(4T_{\text{cry}} (\omega -\epsilon_{\textbf{k}}) +\left(\gamma_0 ^2+(\omega -\epsilon_{\textbf{k}})^2\right) \text{tanh}\left(\frac{\omega -\mu _0}{2T_{\text{cry}}}\right)\right)}{ T_{\text{cry}}^2
  \left(\gamma_0 ^2+(\omega -\epsilon_{\textbf{k}})^2\right)^2}
\end{align}
\section{Effective temperatures}

From the form of the nonthermal steady state one can calculate different effective temperatures. The on-shell effective tempereature we define as coming from $F(\textbf{k}_F,\epsilon_{\textbf{k}})$ so that we have :
\begin{align}
    & F(\textbf{k}_F,\epsilon_{\textbf{k}})= \dfrac{1}{2T_{\text{cry}}}\left(1-\dfrac{ g^2\nu_0(\Delta T+\nu_0^2/(6T_{\text{cav}})) }{4\pi\gamma_0^2T_{\text{cry}}^2}\right)(\epsilon_{\textbf{k}}-\mu_0)+\text{O}(\epsilon_{\textbf{k}}-\mu_0)^3\\
     &T_{\text{eff}}^{\text{on-shell}}=(1-\alpha)^{-1}T_{\text{cry}}, \quad \alpha=\dfrac{g^2\nu_0(\Delta T+\nu_0^2/(6T_{\text{cav}})) }{4\pi \gamma_0^2T_{\text{cry}}^2}
\end{align}
We can instead define a different effective temperature which can be calculated from the nonthermal distribution function by fixing the momentum to be the Fermi momentum, essentially $\epsilon(\textbf{k})=\mu_0$, but not fixing the frequency dependence, so that we have:
\begin{align}
      & F(\textbf{k}_F,\omega)= \dfrac{1}{2T_{\text{cry}}}\left(1-\dfrac{g^2\nu_0(\Delta T+\nu_0^2/(6T_{\text{cav}})) }{4\pi}\dfrac{(8T_{\text{cry}}^2+\gamma_0^2)}{\gamma_0^4T_{\text{cry}}^2}\right)(\omega-\mu_0)+\text{O}(\omega-\mu_0)^3\\
      & \tilde{T}_{\text{eff}}= (1-\tilde{\alpha})^{-1}T_{\text{cry}}, \quad \tilde{\alpha}= \dfrac{g^2\nu_0(\Delta T+\nu_0^2/(6T_{\text{cav}})) }{4\pi}\dfrac{(8T_{\text{cry}}^2+\gamma_0^2)}{\gamma_0^4T_{\text{cry}}^2}
\end{align}
\section{Approximate solution for non-Markovian bath}

Let us now analyze the solution for a non-Markovian bath, which means we have corrections due to the fact that $\gamma=\gamma(\omega)$ with a frequency dependence. Following the steps from the previous derivation we obtain to first order in frequency and in $\kappa=-\dfrac{g^2\nu_0 \Delta T}{4\pi T_{\text{cry}}}$, now a similar form to before plus an extra term:

\begin{align}
    \Delta F=-\dfrac{g^2\nu_0 \gamma(\omega)}{4\pi A(\omega)}\dfrac{\partial}{\partial \omega}\left(\dfrac{A(\omega)^2}{\gamma(\omega)^2}\left(1-F_0(\textbf{k},\omega)^2-2T_{\text{cav}}\dfrac{\partial}{\partial \omega}F_0\textbf{k},\omega)\right)\right)=-\dfrac{\kappa \gamma(\omega)}{ A(\omega)}\dfrac{\partial}{\partial \omega}\left(\dfrac{A(\omega)^2}{\gamma(\omega)^2}\left(\sech\left(\dfrac{\omega-\mu_0}{2T_{\text{cry}}}\right)\right)^2\right)
\end{align}
\section{Modified Sommerfeld expansion}

The consequences of the nonthermal steady state will generally manifest on the level of observables via a modified Sommerfeld expansion. Let us first recall how the usual Sommerfeld expansion goes and then modify it accordingly. We can derive this from considering the expectation value of a single particle observable $\expval{\hat{O}}$ of the generic form:

\begin{align}
\expval{\int dt \sum_{\textbf{k}}o_{\textbf{k}}(t) \psi_{\textbf{k}}(t)^\dagger \psi_{\textbf{k}}(t) } = \sum_{\textbf{k}} \int dt  \dfrac{o_{\textbf{k}}(t)}{2}\left( 1-i G_{\textbf{k}}^K(t)\right)=  \sum_{\textbf{k}} \int dt \int \dfrac{\dd \omega_1}{2\pi}  \dfrac{o_{\textbf{k}}(\omega_1)}{2}e^{i\omega_1 t}\left(\int  \dfrac{\dd \omega_2}{2\pi} 1-i G_{\textbf{k}}^K(\omega_2)e^{i\omega_2 t}\right),
\end{align}
where we used the identity of the Keldysh green's function $iG^K=1-2\expval{\psi^\dagger \psi}$ so that now we can use the parametrisation in terms of $F(\omega)$ and the retarded and advanced greens functions to get:
\begin{align}
  \expval{\hat{O}}= \dfrac{1}{2}  \sum_{\textbf{k}}\left[ \int \dfrac{\dd \omega}{2\pi} 2\Im{G^R_{\textbf{k}}(\omega)}o_{\textbf{k}}(-\omega)F(\textbf{k},\omega)+o_{\textbf{k}}(0)\right]=\dfrac{1}{2} \sum_{\textbf{k}}\left[-\int \dfrac{\dd \omega}{2\pi} 2 \pi \delta(\omega-\epsilon_{\textbf{k}})o_{\textbf{k}}(-\omega)F(\textbf{k},\omega)+o_{\textbf{k}}(0)\right],
\end{align}
where we assumed that the width of the spectral function $\gamma$ is much smaller than the region where the observable and occupation function vary. To first order this would mean a big electron temperature $T_{\text{cry}}\gg \gamma$. Then we would obtain:
\begin{align}
     \expval{\hat{O}}=  \dfrac{1}{2}\sum_{\textbf{k}} [o_{\textbf{k}}(0)-o_{\textbf{k}}(-\epsilon_{\textbf{k}})F(\textbf{k},\epsilon_{\textbf{k}})]=\dfrac{1}{2} \int \dfrac{\dd^d \textbf{k}}{(2\pi)^d} o_{\textbf{k}} [1-F(\epsilon_{\textbf{k}})]= \dfrac{1}{2} \int \dd \epsilon \ O(\epsilon)\  D(\epsilon) [1-F(\epsilon)]= \int \dd \epsilon \ H(\epsilon) \dfrac{1}{2}[1-F(\epsilon)],
\end{align}
where we assumed also $o_{\textbf{k}}=o_{\textbf{k}}(0)=o_{\textbf{k}}(-\epsilon_{\textbf{k}})=O(\epsilon_{\textbf{k}})$ we also used in the last two equalities spherical coordinates and assuming a spherical dispersion to transform to energy space and defined the density of states as $D(\epsilon)$. We also defined the notation $H(\epsilon)=O(\epsilon)D(\epsilon)$. We obtain then the usual Sommerfeld expansion by identifying $n(\epsilon)=\dfrac{1}{2}(1-F(\epsilon))$ which is the usual definition of the Fermi distribution function in the thermal state. We have thus reached the usual form of the initial integral before the Sommerfeld expansion is performed. In our case we have to be careful since the retarded propagator has a finite width which means now we stop before the delta function is introduced and divide the contribution in two since we write $F=F_0+\Delta F$, where the thermal part $F_0$ will follow the usual Sommerfeld expansion so we neglect it here and study the purely nonthermal effect which means we have a contribution of the form:
\begin{align}
    \Delta O=\sum_{\textbf{k}} \int \dfrac{\dd \omega}{2\pi} \Im{G^R_{\textbf{k}}(\omega)}o_{\textbf{k}}(-\omega)\Delta F(\textbf{k},\omega)=-\sum_{\textbf{k}} o_{\textbf{k}} \int \dfrac{\dd \omega}{2\pi} \dfrac{\gamma(\omega)}{(\omega-\epsilon_{\textbf{k}})^2+\gamma(\omega)^2}\Delta F(\textbf{k},\omega),
\end{align}
where we assumed the observable is purely momentum dependent like energy or density so that we neglect its frequency dependence. Let us call the last expression $\tilde{f}(\epsilon_{\textbf{k}})$ it will be given by:
\begin{align}
    \tilde{f}(\epsilon_{\textbf{k}})=-\int \dfrac{\dd \omega}{2\pi} \dfrac{\gamma(\omega)}{(\omega-\epsilon_{\textbf{k}})^2+\gamma(\omega)^2}\Delta F(\textbf{k},\omega)
    \approx \int \dfrac{\dd \omega}{2\pi} \left\{\kappa\gamma(\omega)\dfrac{\partial}{\partial \omega}\left( \frac{\text{sech}^2\left(\frac{\omega -\mu _0}{
   T_{\text{cry}}}\right)}{
   \left[(\omega-\epsilon_{\textbf{k}})^2+\gamma(\omega)^2\right]^2}\right)\right\},
\end{align}
after integrating by parts we obtain the expression:
\begin{align}
      \tilde{f}(\epsilon_{\textbf{k}})=-\kappa\int_{-\infty}^{\infty} \dfrac{\dd \omega}{2\pi} \frac{\gamma'(\omega)\text{sech}^2\left(\frac{\omega -\mu _0}{
   2T_{\text{cry}}}\right)}{
   \left((\omega-\epsilon_{\textbf{k}})^2+\gamma(\omega)^2\right)^2}\approx-\dfrac{2\kappa T_{\text{cry}}}{\pi} \frac{\gamma'(\mu_0)}{
   \left((\mu_0-\epsilon_{\textbf{k}})^2+\gamma(\mu_0)^2\right)^2}
\end{align}
,where we used the fact that $\text{sech}^2\left(\frac{\omega -\mu _0}{2 T_{\text{cry}}}\right)$ is highly peaked at $\mu_0$ for small enough $T_{\text{cry}}$ so that we can to first order evaluate the rest of the integral at the peaked value and use the fact that $\int_{-\infty}^{\infty}\dd \text{x} \ \text{sech}^2\left(\frac{\text{x}}{
   2T_{\text{cry}}}\right)=4T_{\text{cry}}$.
Let us now use the density of states just as before to rewrite the momentum integral in terms of an integration to obtain now:
\begin{align}
   \Delta O= \int_{-\infty}^{\infty} \dd \epsilon \ H(\epsilon) \tilde{f}(\epsilon)\approx H(\mu_0)\int_{-\infty}^{\infty} \dd \epsilon \tilde{f}(\epsilon)=\dfrac{ g^2\nu_0 \Delta T}{4\pi T_{\text{cry}}}\dfrac{2 T_{\text{cry}}}{\pi}\dfrac{\pi}{2\gamma_0^3}H(\mu_0)\gamma'(\mu_0)
\end{align}
where we again used the fact that $\tilde{f}(\epsilon)$ is peaked at $\mu_0$ for $\mu_0\gg \gamma(\mu_0)$ so that we can expand the integrand to lowest order and perform the integration, we also substituted back the value of $\kappa$. In conclusion the nonthermal steady state gives an anomalous contribution to the Sommerfeld expansion given by:
\begin{align}
    \Delta O= \dfrac{ g^2\nu_0 \Delta T}{4\pi\gamma_0^3} O(\mu_0)D(\mu_0)\gamma'(\mu_0)
\end{align}

,with $\Delta T=T_{\text{cav}}-T_{\text{cry}}$ for absorbing the minus sign and defined $\gamma(\mu_0)=\gamma_0$.

\subsection{Time dependent observables}
For a time dependent observable the calculation is very similar and in this case we get a contribution even for the case of constant damping $\gamma(\omega)=\gamma_0=\text{cte}$. We start from the frequency dependent expression for the change in the observable, we denote now $O(\omega)$ the frequency dependent observable, the change is then:
\begin{align}
    \Delta O=-\int_{-\infty}^{\infty} \dd \epsilon \int_{-\infty}^{\infty}\dfrac{ \dd \omega }{2 \pi} \dfrac{O(\omega)\gamma_0}{(\omega-\epsilon)^2+\gamma_0^2}\Delta F(\epsilon,\omega) D(\epsilon)=\kappa \int_{-\infty}^{\infty} \dd \epsilon \int_{-\infty}^{\infty}\dfrac{ \dd \omega }{2 \pi} O(\omega) \dfrac{\partial}{\partial \omega}\left(\dfrac{\gamma_0\sech^2(\frac{\omega-\mu_0}{T_{\text{cry}}})}{((\omega-\epsilon)^2+\gamma_0^2)^2}\right) D(\epsilon),
\end{align}
where we used the density of states again $D(\epsilon)$ to reexpress the momentum integral in terms of an energy integral. We integrate now by parts just as in the previous case and use the strongly peak nature of the hyperbolic secant to get:
\begin{align}
    \Delta O=-\kappa \int_{-\infty}^{\infty} \dd \epsilon \int_{-\infty}^{\infty}\dfrac{ \dd \omega }{2 \pi}  D(\epsilon)O'(\omega)\dfrac{\gamma_0\sech^2(\frac{\omega-\mu_0}{T_{\text{cry}}})}{((\omega-\epsilon)^2+\gamma_0^2)^2}\approx  -\int_{-\infty}^{\infty} \dd \epsilon \dfrac{\kappa \gamma_0 4T_{\text{cry}}}{2\pi}D(\epsilon)O'(\omega)\dfrac{1}{((\omega-\epsilon)^2+\gamma_0^2)^2}
\end{align}
Finally just as before we take the limit $\mu_0\gg \gamma_0$ so as to obtain now the integrand evaluated at the chemical potential and the integral of the squared lorentzian to give the same factor as before, resulting in:
\begin{equation}
    \Delta O\approx -\dfrac{\kappa \gamma_0 4T_{\text{cry}}}{2\pi}D(\mu_0)\eval{O'(\omega)}_{\omega=\mu_0} \dfrac{\pi}{2\gamma_0^3}= \dfrac{ g^2\nu_0 \Delta T}{4\pi T_{\text{cry}}}\dfrac{\gamma_0 4T_{\text{cry}}}{2\pi} \dfrac{\pi}{2\gamma_0^3}D(\mu_0)\eval{O'(\omega)}_{\omega=\mu_0}.
\end{equation}
Simplifying we obtain now the analogous result for a time dependent observable where the modified contribution to the Sommerfel expansion comes from the frequency dependence of the observable:
\begin{align}
        \Delta O= \dfrac{ g^2\nu_0 \Delta T}{4\pi\gamma_0^2}D(\mu_0)\eval{O'(\omega)}_{\omega=\mu_0}
\end{align}

\end{document}